%% file: TOSEM_MAP4CS.tex
\documentclass[manuscript]{acmart}

\AtBeginDocument{%
  \providecommand\BibTeX{{\rmfamily B\kern-.05em\textsc{i\kern-.025em b}\kern-.08em\TeX}}}

\setcopyright{acmlicensed}
\copyrightyear{2026}
\acmYear{2026}
\acmDOI{XXXXXXX.XXXXXXX}

\acmJournal{TOSEM}
\acmVolume{XX}
\acmNumber{X}
\acmArticle{XXX}
\acmMonth{6}

\input{macro}

\makeatletter
\def\acm@nullfootnoterule{\kern-3\p@\hrule\@width\z@\kern2.6\p@}
\AtBeginDocument{%
  \expandafter\def\csname MFL@join\number\footinscopyrightpermission\endcsname#1#2{%
    \MFL@joinrule{2}{\acm@nullfootnoterule}{}{#1}{#2}}%
  \skip\footinscopyrightpermission=\z@
}
\makeatother

\begin{document}

%% --------------------------------------------------------------------
%% Title
%% --------------------------------------------------------------------
\title{\ourapproach: A Multi-dimensional Data Pruning Framework for
       Efficient Code Retriever Fine-tuning}

%% --------------------------------------------------------------------
%% Authors and affiliations (compact: one affiliation per author;
%% the Zhuhai Key Laboratory listing is omitted to keep the bibstrip
%% concise --- it can be added back via a second \affiliation block
%% if needed for funding/acknowledgement reasons.)
%% --------------------------------------------------------------------
\author{Yuxuan Chen}
\affiliation{%
  \institution{School of Software Engineering, Sun Yat-sen University}
  \city{Zhuhai}\country{China}}
\email{chenyx677@mail2.sysu.edu.cn}

\author{Mingwei Liu}
\authornote{Corresponding author.}
\affiliation{%
  \institution{School of Software Engineering, Sun Yat-sen University}
  \city{Zhuhai}\country{China}}
\email{liumw26@mail.sysu.edu.cn}

\author{Guangsheng Ou}
\affiliation{%
  \institution{School of Software Engineering, Sun Yat-sen University}
  \city{Zhuhai}\country{China}}
\email{ougsh3@mail2.sysu.edu.cn}

\author{Zekai Zhang}
\affiliation{%
  \institution{School of Software Engineering, Sun Yat-sen University}
  \city{Zhuhai}\country{China}}
\email{zhangzk27@mail2.sysu.edu.cn}

\author{Zike Li}
\affiliation{%
  \institution{School of Software Engineering, Sun Yat-sen University}
  \city{Zhuhai}\country{China}}
\email{lizk8@mail2.sysu.edu.cn}

\author{Yanlin Wang}
\affiliation{%
  \institution{School of Software Engineering, Sun Yat-sen University}
  \city{Zhuhai}\country{China}}
\email{yanlin-wang@outlook.com}

\author{Peilin Zheng}
\affiliation{%
  \institution{School of Software Engineering, Sun Yat-sen University}
  \city{Zhuhai}\country{China}}
\email{zhengplin@mail.sysu.edu.cn}

%% Short author list shown in running headers
\renewcommand{\shortauthors}{Chen et al.}

%% --------------------------------------------------------------------
%% Abstract
%% --------------------------------------------------------------------
\begin{abstract}
\input{sections/0_abstract}

\end{abstract}

%% --------------------------------------------------------------------
%% CCS concepts (placeholder --- generate the real codes at
%%   https://dl.acm.org/ccs and replace the block below before camera-ready)
%% --------------------------------------------------------------------


%% --------------------------------------------------------------------
%% Keywords (kept identical to TSE submission per author preference)
%% --------------------------------------------------------------------

%% Optional: received dates (uncomment and fill once known)
% \received{XX Month 2026}
% \received[revised]{XX Month 2026}
% \received[accepted]{XX Month 2026}

\maketitle

%% --------------------------------------------------------------------
%% Body
%% --------------------------------------------------------------------
\input{sections/1_introduction}

\input{sections/2_background}

\input{sections/3_Approach}
\input{sections/4_Experiment}
\input{sections/5_Result/5_Result}

\input{sections/6_Threats}

\input{sections/7_Related}

\input{sections/8_conclusion}

\input{sections/9_data_availability}

%% --------------------------------------------------------------------
%% References (ACM Reference Format with the existing ref.bib)
%% --------------------------------------------------------------------
\bibliographystyle{ACM-Reference-Format}
\bibliography{ref}

\end{document}

%% file: macro.tex
\usepackage{amssymb}
\usepackage{amsfonts}

\usepackage{algorithm}
\usepackage{algpseudocode}

\usepackage{makecell}
\usepackage{tabularx}
\usepackage{multirow}
\usepackage{array}
\usepackage{threeparttable}
\usepackage{colortbl}
\usepackage{hhline}
\usepackage{diagbox}

\usepackage{enumitem}
\usepackage{xspace}
\usepackage{soul}
\usepackage{pifont}

\usepackage{float}
\usepackage{adjustbox}
\usepackage{mdframed}
\usepackage{tcolorbox}
\tcbuselibrary{most, skins, breakable, theorems}

\usepackage{tikz}
\usepackage{pgfplots}
\pgfplotsset{compat=1.16}
\usetikzlibrary{pgfplots.statistics}

\newcolumntype{C}{>{\centering\arraybackslash}X}

\newcommand{\rev}[1]{\textcolor{black}{#1}}

\newcommand{\parabf}[1]{\noindent\textbf{#1}}

\definecolor{ggray}{HTML}{eff0f0}
\definecolor{gggray}{HTML}{E8E8E8}
\definecolor{ggggray}{HTML}{BEBEBE}
\definecolor{myyellow}{HTML}{FFF2CC}

\newcommand{\ourapproach}{MAP4CS\xspace}
\newcommand{\queryfourcode}{Query4Code\xspace}
\newcommand{\texttwosql}{Synthetic\xspace Text2SQL\xspace}
\newcommand{\csn}{CodeSearchNet\xspace}
\newcommand{\cosqa}{CoSQA\xspace}
\newcommand{\apps}{APPS\xspace}
\newcommand{\gte}{GTE$_{base}$\xspace}
\newcommand{\unixcoder}{UniXcoder\xspace}

\newcounter{finding}
\newcommand{\finding}[1]{\refstepcounter{finding}
    \begin{mdframed}[linecolor=gray!25,roundcorner=12pt,backgroundcolor=myyellow!30,linewidth=3pt,innerleftmargin=2pt, leftmargin=0cm,rightmargin=0cm,topline=false,bottomline=false,rightline=false]
        \textbf{Finding \arabic{finding}:} #1
    \end{mdframed}
}

\newcommand{\boxmargin}{1mm}
\newtcolorbox{myboxa}[2][]{
    colback=gray!10!white,
    colframe=black, enhanced,
    attach boxed title to top left={yshift=-2mm,xshift=5mm},
    title=#2,#1
}
\newtcolorbox{myboxb}[2][]{
    boxsep=3pt,
    left=\boxmargin, right=\boxmargin, top=\boxmargin, bottom=\boxmargin,
    title={#2},#1
}
\newtcolorbox{myboxc}{
    colback=gray!15!white,
    arc=0pt, outer arc=0pt,
    boxsep=0pt, left=3pt, right=0pt, top=0pt, bottom=0pt,
    leftrule=3pt, bottomrule=0pt, toprule=0pt, rightrule=0pt,
    left=\boxmargin, right=\boxmargin, top=\boxmargin, bottom=\boxmargin
}
\newtcolorbox{myboxd}{
    colback=gray!10,
    colframe=black,
    width=\columnwidth,
    arc=1mm, auto outer arc,
    boxrule=0.5pt,
}

%% file: sections/0_abstract.tex
Retrieval-Augmented Generation (RAG) has become a cornerstone in software engineering for enhancing Large Language Models (LLMs) with domain-specific knowledge. However, adapting retrievers to evolving code repositories remains challenging due to the noise and redundancy inherent in massive code corpora. Standard fine-tuning on the full corpus is computationally expensive and often leads to sub-optimal performance due to negative transfer from low-quality samples. 
Conversely, simple random sampling fails to guarantee data representativeness.

To address these challenges, we propose \ourapproach (Multi-dimensional Awareness Pruning for Code Search), an adaptive data pruning framework.
\ourapproach identifies a small, high-quality core subset by integrating syntactic structure, semantic diversity, and distributional
representation, followed by a rigorous rule-based filtering pipeline. 
Extensive experiments on two large-scale datasets demonstrate that \ourapproach consistently outperforms random sampling baselines using only \textbf{5\%} of the training data. Remarkably, it achieves performance comparable to, or even superior to, fine-tuning on the full dataset, validating the ``less is more'' hypothesis in data-centric AI. Furthermore, linguistic analysis reveals an adaptive optimization mechanism: \ourapproach automatically functions as a de-duplicator for redundant corpora and a denoiser for chaotic ones, constructing a training corpus that is both lexically diverse and information-dense.

%% file: sections/1_introduction.tex
\section{Introduction}

The advent of Large Language Models (LLMs) has fundamentally reshaped the landscape of software engineering, empowering automated tools for code completion, summarization, and synthesis~\cite{DBLP:journals/corr/abs-2311-10372,DBLP:journals/ese/ZhengNZCCGWW25}. 
Despite their impressive capabilities, off-the-shelf LLMs often struggle with domain-specific software repositories, particularly when dealing with private enterprise libraries, evolving APIs, or legacy systems that were not part of their pre-training data~\cite{DBLP:conf/emnlp/ZhangCZKLZMLC23}. 
To mitigate limitations such as hallucinations and knowledge cutoffs, Retrieval-Augmented Generation (RAG) has emerged as a pivotal paradigm~\cite{DBLP:conf/nips/LewisPPPKGKLYR020, DBLP:conf/emnlp/ParvezACRC21}. By retrieving relevant code snippets from an external knowledge base to serve as context, RAG significantly enhances the accuracy and reliability of generated code~\cite{DBLP:conf/nips/LewisPPPKGKLYR020, DBLP:conf/emnlp/ZhangCZKLZMLC23}. 
Consequently, the performance of the neural code retriever becomes the bottleneck of the entire pipeline~\cite{DBLP:journals/tmlr/IzacardCHRBJG22}.

Regrettably, generic code retrievers often exhibit sub-optimal performance in these specialized software domains due to a lack of exposure to relevant proprietary knowledge during pre-training. To adapt general-purpose retrievers to these domains, fine-tuning on domain-specific corpora is essential~\cite{DBLP:journals/corr/abs-2410-22240}. This process typically involves extracting pairs of ``Natural Language Query'' and ``Code Snippet'', denoted as $(Q, C)$, from the target repositories~\cite{DBLP:journals/corr/abs-1909-09436}. 
The model is then fine-tuned on these sample pairs to align the embedding space of the retriever with the specific coding standards and business logic of the target domain~\cite{DBLP:conf/icse/GuZ018}.

A common practice in the industry is to fine-tune models on the full available corpus to maximize knowledge coverage~\cite{DBLP:journals/corr/abs-2001-08361, DBLP:journals/corr/abs-2402-19173}. However, blindly utilizing all data presents a critical trade-off between efficiency and effectiveness.
\textbf{First, the ``Noise Toxicity'' Problem:} Real-world code repositories are fraught with noise, such as redundant boilerplate code, low-quality auto-generated comments, and deprecated functions~\cite{DBLP:conf/oopsla/Allamanis19, DBLP:journals/tmlr/KocetkovLALMJMF23}. Training on such data can lead to \textit{negative transfer}, where the retriever overfits to superficial lexical overlaps rather than learning robust semantic representations~\cite{DBLP:conf/cvpr/ChuangHWVJ0JS22}.
\textbf{Second, the Computational Barrier:} The sheer scale of industrial repositories makes full-corpus fine-tuning computationally prohibitive and environmentally costly~\cite{DBLP:conf/acl/StrubellGM19}, creating a significant barrier for frequent updates in continuous integration (CI) environments.

To reduce training costs, a straightforward alternative is Random Sampling~\cite{DBLP:conf/nips/SorscherGSGM22, DBLP:conf/icml/MirzasoleimanBL20}. While this decreases fine-tuning time, this content-agnostic strategy is flawed for code search tasks. It often leads to the Loss of Long-Tail Knowledge, discarding crucial but infrequent domain-specific APIs~\cite{DBLP:conf/icse/HindleBSGD12}, while simultaneously suffering from the Retention of High-Entropy Noise, preserving misaligned pairs that confuse the model's decision boundaries~\cite{DBLP:journals/cacm/ZhangBHRV21}.
Thus, neither full-data training nor random sampling offers a viable path for efficient and effective domain adaptation.

This dilemma poses a central research question: \textit{Is it possible to intelligently identify a small, high-quality subset of data that trains a better code retriever than the full corpus?} 
While data pruning has shown promise in general NLP, directly applying these text-centric methods to code retrieval is insufficient. As we will discuss in Section~\ref{sec:motivation}, generic filtering often ignores the unique structural dependencies of code, while traditional software engineering approaches tend to focus on defensive filtering to remove noise, rather than selecting samples with high syntactic and semantic value.

To address these challenges, we propose \textbf{\ourapproach} (\textbf{M}ulti-dimensional \textbf{A}wareness \textbf{P}runing for \textbf{C}ode \textbf{S}earch), an adaptive data pruning framework designed specifically for efficient retriever fine-tuning.
Unlike rigid filtering rules or general-purpose scorers, \ourapproach treats data selection as a multi-objective optimization problem within local semantic contexts.
It operates through a rigorous coarse-to-fine pipeline: 
First, it structures the chaotic code corpus using Stream-Based Semantic Clustering to localize diversity and isolate noisy regions. 
Second, it evaluates samples via a Multi-Metric Scoring mechanism, integrating \textit{Syntactic Structural Diversity}, \textit{Semantic Alignment Consistency}, and \textit{Distributional Representativeness}. 
Finally, a cascade filtering and sampling strategy selects the core subset. 
This multi-dimensional awareness allows \ourapproach to adaptively function as a ``de-duplicator'' for redundant corpora, a ``denoiser'' for chaotic datasets, and a ``structure-preserver'' for complex logic.

To evaluate \ourapproach, we conduct extensive experiments on two large-scale datasets, CSN-Java and Query4Code, using both code-specialized (UniXcoder~\cite{DBLP:conf/acl/GuoLDW0022}) and general-purpose (\gte~\cite{DBLP:conf/emnlp/ZhangZLXDTLYXHZ24}) , and \rev{foundation-model-based (Qwen3-Embedding~\cite{qwen3embedding2025})} dense encoders, \rev{alongside BM25 as a lexical baseline}. The results demonstrate that using only \textbf{5\%} of the training data, \ourapproach consistently outperforms random sampling baselines \rev{ across all dense encoders} and, remarkably, achieves comparable or even superior performance to fine-tuning on the full dataset. For instance, on the CSN-Java dataset, \ourapproach improves the retrieval MRR by 31.9\% over random sampling \rev{ with UniXcoder and by 45.5\% with Qwen3-Embedding}, highlighting the ``less is more'' phenomenon in data-centric AI.

In summary, this paper makes the following contributions:
\begin{itemize}
    % \item We introduce \textbf{\ourapproach}, a novel multi-dimensional data pruning framework tailored for code search. It integrates syntactic structural diversity, semantic alignment consistency, and distributional representativeness to identify high-value code-query pairs for code retriever fine-tuning.
    \item We introduce \textbf{\ourapproach}, a multi-dimensional pruning framework tailored for code search that selects high-value samples by integrating syntactic structure, semantic alignment, and distributional density.
    \item We propose an adaptive selection mechanism that effectively resolves the data scale-quality trade-off by filtering high-entropy noise while maximizing sample diversity.
    \item Extensive experiments demonstrate that \ourapproach reduces training costs by over 90\% while achieving performance comparable to or superior to full-dataset fine-tuning, validating the ``less is more'' hypothesis.  
    % \item We present an adaptive optimization mechanism: \ourapproach improves data quality by maximizing structural and semantic diversity while filtering out high-entropy noise, effectively solving the trade-off between data scale and quality.
    % \item Comprehensive experiments demonstrate that \ourapproach significantly reduces training costs (by over 90\%) while maintaining or enhancing retrieval performance across multiple benchmarks and model architectures, validating the ``less is more'' hypothesis in code search tasks.
\end{itemize}

%% file: sections/2_background.tex
\section{Background and Motivation}
\label{sec:motivation}

\begin{figure*}[t]
    \centering
    % Placeholder for the figure
    \includegraphics[width=1.0\textwidth]{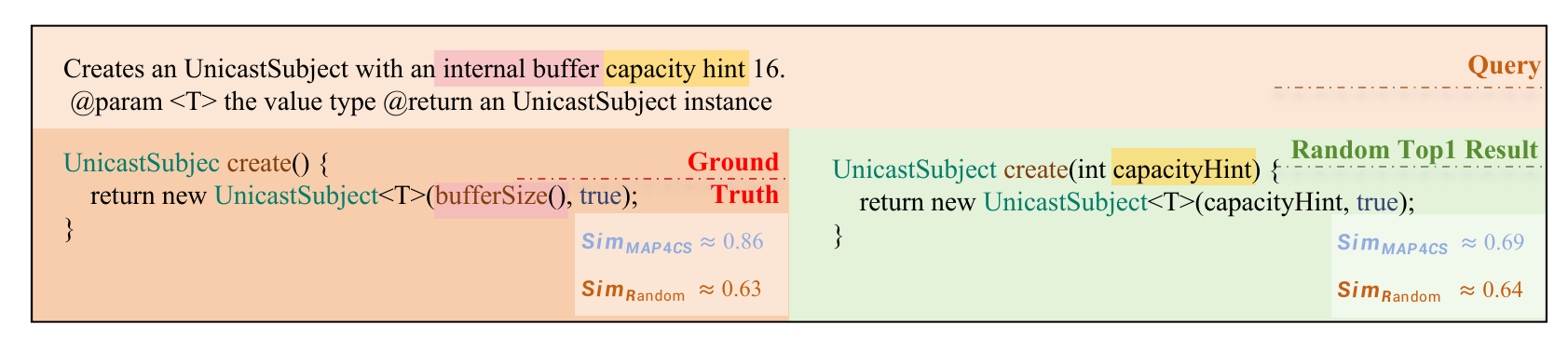}
    % \framebox{\parbox{0.95\columnwidth}{\centering \vspace{1.5cm} \textbf{[Image Placeholder: Semantic Gap Case Study]} \\ \small (Please replace with: pictures/3_Approach/Approach_case_study_17.pdf) \vspace{1.5cm}}}
    \caption{\textbf{Motivation Example 1 (Semantic Gap):} The baseline model is misled by the explicit keyword ``capacity hint'' in a distractor method. In contrast, training with diverse semantic strata helps identify the Ground Truth which handles this logic implicitly.}
    \label{fig:mot_semantic}
    % \vspace{-15pt}
\end{figure*}
% \vspace{-15pt}

In this section, we analyze the limitations of existing data selection methodologies when applied to code retrieval and present motivational examples that necessitate a multi-dimensional pruning approach.

\subsection{The Landscape of Data Pruning}

In the general domain, sophisticated selection strategies have matured into two dominant paradigms: \textit{metric-based heuristics} and \textit{learning-based optimization}. The former relies on statistical signals—such as perplexity or error margins—to filter data, as demonstrated by the ``Textbooks Are All You Need'' hypothesis~\cite{DBLP:journals/corr/abs-2306-11644} and LIMA~\cite{DBLP:conf/nips/ZhouLX0SMMEYYZG23}. The latter leverages training dynamics, employing techniques like Influence Functions~\cite{DBLP:conf/icml/KohL17}, Core-set Selection~\cite{DBLP:conf/icml/MirzasoleimanBL20}, or tracking Forgetting Events~\cite{DBLP:conf/iclr/TonevaSCTBG19} to mathematically quantify the contribution of each sample. Even in instruction tuning, frameworks like AlpaGasus~\cite{DBLP:conf/iclr/ChenLYWGYTS0HJ24} utilize teacher models to score sample complexity.

However, a direct transposition of these text-centric or image-centric algorithms to the domain of code retrieval is fundamentally insufficient. Existing approaches generally fall into two categories, neither of which fully addresses the unique duality of source code~\cite{DBLP:journals/corr/abs-2408-04116} (i.e., natural language intent vs. programming logic).

\textbf{General Data Filtering Ignorance of Structure.} 
General data selection methods typically treat data as flat sequences or continuous vector spaces~\cite{DBLP:conf/nips/ZhouLX0SMMEYYZG23}. When applied to code, metrics like language model perplexity focus solely on the probability of token sequences~\cite{DBLP:journals/corr/abs-2306-11644}. 
This approach ignores the rigid syntactic structures (e.g., Abstract Syntax Trees) and execution logic that define programming languages~\cite{DBLP:conf/oopsla/Allamanis19,DBLP:conf/iclr/GuoRLFT0ZDSFTDC21}. 
A code snippet might be statistically predictable (low perplexity) yet syntactically trivial or functionally irrelevant to the retrieval query. By failing to account for structural integrity, these methods risk selecting ``fluent'' but logically empty boilerplate code, leading to sub-optimal representation learning~\cite{DBLP:conf/icse/ShiWGDZHZS23}.

\textbf{Defensive Nature of Code Cleaning.}
Within the software engineering community, data management efforts have predominantly focused on sanitization rather than valuation. Standard practices involve filtering explicit noise, such as removing near-duplicates (Deduplication)~\cite{DBLP:conf/oopsla/Allamanis19} or discarding snippets with parse errors~\cite{DBLP:journals/tmlr/KocetkovLALMJMF23}. While these ``cleaning'' steps are necessary prerequisites, they are essentially defensive: they eliminate the ``worst'' data but lack the intelligence to actively identify the ``best'' data. They do not assess whether a sample contributes to bridging the \textbf{semantic gap} between natural language queries and code implementations, nor do they consider the distributional redundancy that persists even after deduplication.

Consequently, current methodologies leave us in a compromise: we must choose between general pruning methods that misunderstand code structure, or traditional cleaning methods that lack selection intelligence. This gap underscores the need for a specialized framework tailored to the multi-dimensional nature of code.

\subsection{Motivation}

% To illustrate the limitations of naive data selection strategies (e.g., random sampling) and the necessity of structural and semantic awareness, we analyze two typical failure modes in code retrieval where models trained on randomly sampled data fail to align user intent with code implementation.
To illustrate the limitations of naive data selection strategies and the necessity of structural and semantic awareness, we analyze two typical failure modes in code retrieval where models trained on randomly sampled data fail to align user intent with code implementation.

\textbf{(1) Implicit Semantics vs. Lexical Overlap.} 
As shown in Figure \ref{fig:mot_semantic}, queries often rely on implicit domain knowledge rather than explicit keywords. For a query requesting an ``internal buffer capacity hint'', the baseline model trained on randomly sampled data is distracted by a superficial keyword match in an incorrect method (\texttt{capacityHint}), missing the ground truth which handles capacity implicitly. This failure highlights the necessity of \textbf{Semantic Alignment Consistency}—we select ``hard positives'' during training to teach the model to look beyond surface-level lexical overlaps.

\begin{figure*}[t]
    \centering
    % Placeholder for the figure
    % \framebox{\parbox{0.95\columnwidth}{\centering \vspace{1.5cm} \textbf{[Image Placeholder: Structural Logic Case Study]} \\ \small (Please replace with: pictures/3_Approach/Approach_case_study_27.pdf) \vspace{1.5cm}}}
    \includegraphics[width=1.0\textwidth]{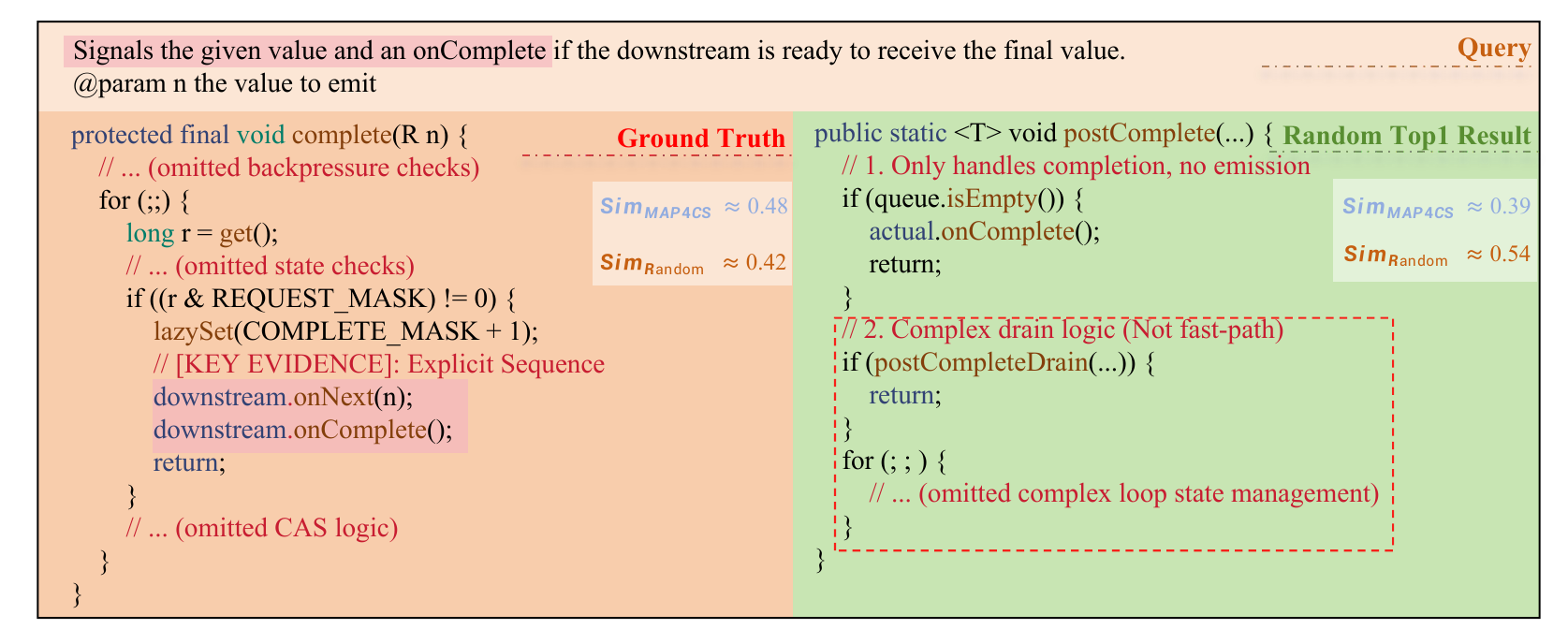}
    \caption{\textbf{Motivation Example 2 (Structural Alignment):} The query implies a specific sequential logic (``onNext'' then ``onComplete''). Without structural awareness, the baseline overfits to domain terms in complex boilerplate, missing the correct control flow found by AST-aware pruning.}
    \label{fig:mot_structural}
    % \vspace{-15pt}
\end{figure*}
% \vspace{-15pt}

\textbf{(2) Structural Logic Alignment.} 
Figure \ref{fig:mot_structural} demonstrates the challenge of distinguishing specific execution logic from generic boilerplate. The query implies a specific sequence of actions (``Signal value AND onComplete''). Lacking structural guidance, the baseline overfits to shared domain terms (e.g., ``Complete'', ``Queue'') in a complex utility method. This underscores the need for \textbf{Syntactic Structural Diversity}. By incorporating AST-based pruning, we ensure the training data covers diverse control flows, enabling the model to align natural language action sequences with the actual code execution logic.

These failure modes highlight that effective data pruning must transcend simple statistical sampling; \ourapproach addresses this by explicitly optimizing for both semantic alignment and structural diversity, ensuring the retention of high-value samples that are critical for generalization but often discarded by naive baselines.

%% file: sections/3_Approach.tex
% \vspace{-15pt}
\section{Approach}
\label{sec:app}

\begin{figure*}[t]
	\centering
    %\vspace{-6mm}
	\includegraphics[width=0.95\textwidth]{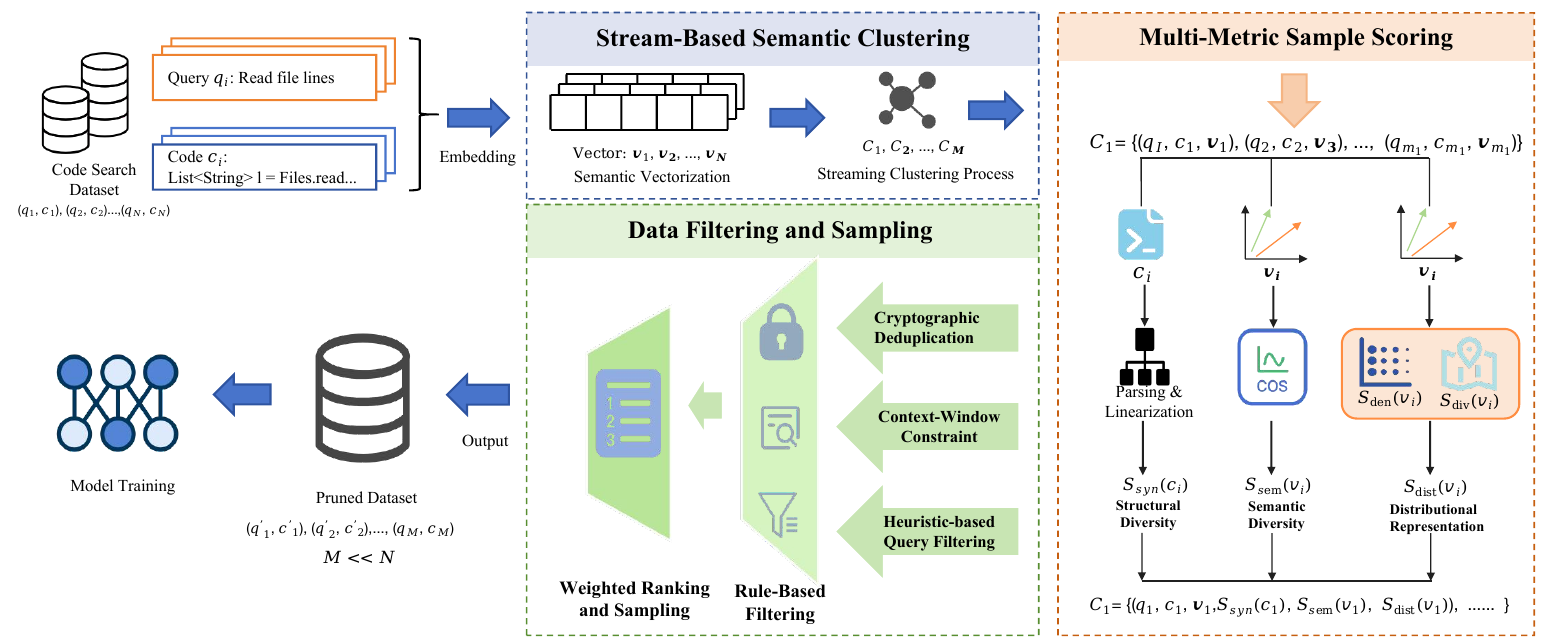}
    % \vspace{-20pt}
    \caption{Overview of \ourapproach}
    \label{fig:overview}
    % \vspace{-15pt}
    
\end{figure*}

In this section, we formulate the data pruning problem and present \textbf{\ourapproach} (Multi-dimensional Awareness Pruning for Code Search). 
As illustrated in Figure \ref{fig:overview}, \ourapproach operates through a rigorous, coarse-to-fine pipeline: 
(1) \textbf{Stream-Based Semantic Clustering}, which partitions the streaming corpus into local contexts; 
(2) \textbf{Multi-Metric Sample Scoring}, which evaluates samples based on structure, semantics, and distribution; and 
(3) \textbf{Data Filtering and Sampling}, which executes the final data distillation under a strict budget.

\subsection{Problem Formulation}
Let $\mathcal{D} = \{(q_i, c_i)\}_{i=1}^N$ denote a large-scale, potentially noisy code corpus containing $N$ training pairs, where $q_i$ represents its corresponding natural language query and $c_i$ represents a code snippet. The goal of code retrieval is to learn a scoring function $f_\theta(q, c)$ that maximizes the similarity between relevant pairs.

The data pruning problem can be formalized as finding an optimal subset $\mathcal{S} \subset \mathcal{D}$ that satisfies a strict budget constraint $|\mathcal{S}| \le B$ (where $B \ll N$, e.g., $B = 0.05N$), such that the model trained on $\mathcal{S}$ achieves maximal performance. Formally, we aim to solve:
\begin{equation}
    \mathcal{S}^* = \mathop{\arg\max}_{\mathcal{S} \subset \mathcal{D}, |\mathcal{S}| \le B} \mathbb{E}_{(q, c) \sim \mathcal{D}_{val}} \left[ \mathcal{M} \left( f_{\theta_{\mathcal{S}}}(q, c) \right) \right]
\end{equation}
where $\theta_{\mathcal{S}}$ denotes the parameters of the model fine-tuned on subset $\mathcal{S}$, and $\mathcal{M}(\cdot)$ is the evaluation metric (e.g., MRR) on a validation set.

% \vspace{-15pt}

\subsection{Stream-Based Semantic Clustering}

The original code corpus usually exhibits a long-tailed distribution, with mainstream functions (such as standard I/O operations) often obscuring the logic specific to certain domains. To preserve the intrinsic semantic diversity of the data, we first structure the data into local semantic contexts.

\begin{algorithm}[t]
\caption{Single-Pass Clustering with Dynamic Centroids}
\label{alg:single_pass}
\begin{algorithmic}[1]
\Require Embeddings $\mathcal{E} = \{\mathbf{v}_1, \dots, \mathbf{v}_N\}$, Threshold $\tau$
\Ensure Cluster Labels $\mathcal{L}$, Centroids $\mathcal{C}$
\State Initialize $\mathcal{C} \leftarrow \emptyset$, Counts $\mathcal{N} \leftarrow \emptyset$
\For{$i = 1$ to $N$}
    \State $\mathbf{v}_i \leftarrow \mathcal{E}[i]$
    \If{$\mathcal{C}$ is empty}
        \State Create new cluster $k_{new} \leftarrow 0$
        \State $\mathcal{C}[k_{new}] \leftarrow \mathbf{v}_i$; $\mathcal{N}[k_{new}] \leftarrow 1$
        \State $\mathcal{L}[i] \leftarrow k_{new}$
        \State \textbf{continue}
    \EndIf
    \State Calculate similarities to all centroids
    \State $k^* \leftarrow \arg\max_{j} (\cos(\mathbf{v}_i, \mathbf{c}_j))$ \Comment{Find nearest centroid}
    \State $sim_{max} \leftarrow \max_{j} (\cos(\mathbf{v}_i, \mathbf{c}_j))$
    \If{$sim_{max} \ge \tau$}
        \State $\mathcal{L}[i] \leftarrow k^*$
        \State $n \leftarrow \mathcal{N}[k^*]$
        \State $\mathbf{c}_{k^*} \leftarrow \frac{\mathbf{c}_{k^*} \cdot n + \mathbf{v}_i}{n + 1}$ \Comment{Dynamic Update}
        \State $\mathcal{N}[k^*] \leftarrow n + 1$
    \Else
        \State Create new cluster $k_{new} \leftarrow |\mathcal{C}|$
        \State $\mathcal{C}[k_{new}] \leftarrow \mathbf{v}_i$; $\mathcal{N}[k_{new}] \leftarrow 1$
        \State $\mathcal{L}[i] \leftarrow k_{new}$
    \EndIf
\EndFor
\State \Return $\mathcal{L}$
\end{algorithmic}
\end{algorithm}

\subsubsection{Clustering Algorithm Selection}
\label{sec:clustering}

Standard clustering algorithms pose significant challenges for massive code corpora. 
\textbf{(1) Unknown $K$:} $K$-Means requires a predefined number of clusters ($K$), which is impossible to estimate accurately for evolving, streaming code data. 
\textbf{(2) Noise Sensitivity:} Density-based methods (e.g., HDBSCAN), while effective for shape detection, often classify distinct long-tail functionalities as noise (outliers) in high-dimensional spaces, resulting in data loss.
\textbf{(3) Efficiency:} With $O(N^2)$ complexity, hierarchical methods are computationally prohibitive for datasets with millions of samples.

To address these limitations, we employ the \textbf{Single-Pass Clustering} algorithm~\cite{DBLP:journals/kais/NguyenWN15}. It processes data in a streaming fashion with $O(N)$ complexity and dynamically discovers cluster structures without a fixed $K$.  The core mechanism involves dynamic centroid updates to adapt to concept drift in the data stream. 
The implementation of Single-Pass Clustering is shown in Algorithm \ref{alg:single_pass}.
% The specific implementation steps of Single-Pass Clustering are shown in Algorithm \ref{alg:single_pass}.
% Single-Pass Clustering具体实现步骤如Algorithm \ref{alg:single_pass}所示

\subsubsection{Clustering Process}

The implementation of our clustering phase involves three key steps: vectorization, streaming clustering, and threshold determination.

\noindent\textbf{Semantic Vectorization.} Before formal clustering, we utilize a pre-trained encoder (e.g., UniXcoder) to map each code-query pair $(q_i, c_i)$ into a dense semantic vector $\mathbf{v}_i$. This step ensures that subsequent operations are performed in a high-dimensional semantic space capturing deep functional relationships.

\noindent\textbf{Streaming Clustering Process.} The core mechanism involves dynamic centroid updates to adapt to the streaming nature of the data. As detailed in Algorithm \ref{alg:single_pass}, for each incoming sample $\mathbf{v}_i$, we calculate its cosine similarity with existing cluster centroids $\mathcal{C}$. If the maximum similarity exceeds a threshold $\tau$, the sample is assigned to the best-matching cluster, and the centroid is updated via a weighted moving average (Line 9). Otherwise, a new cluster is initialized.

\noindent\textbf{Threshold Determination.} 
Based on preliminary experiments, we set the similarity threshold at $\tau = 0.55$.
% To ensure optimal granularity, we conduct extensive preliminary experiments and empirically fix the similarity threshold at $\tau = 0.55$. 
This threshold strikes a balance between \textit{intra-cluster cohesion} and \textit{semantic coverage}: it ensures that elements within each cluster are semantically relevant (high cohesion) while preventing the formation of excessive singleton clusters (over-fragmentation) or overly broad categories.

\subsection{Multi-Metric Sample Scoring}

Within each local semantic cluster, simply selecting the "most similar" samples leads to homogeneity. \ourapproach{} evaluates samples across three independence metrics to ensure diversity and representativeness: Structural Diversity ($S_{syn}$), Semantic Consistency ($S_{sem}$), and Distributional Representation ($S_{dist}$).

\subsubsection{\textbf{Structural Diversity ($S_{syn}$)}}

% Neural retrieval models often tend to overfit surface-level lexical features (e.g., specific variable naming styles) 
Neural retrieval models often tend to overfit surface-level lexical features 
rather than learning deep programming logic structures\cite{DBLP:conf/icse/ShiWGDZHZS23}. To enable the model to perceive different syntactic structures and enhance its generalization ability, the training data need include diverse syntactic implementations of the same functionality.

We introduce an Abstract Syntax Tree (AST)-based method to quantify structural diversity. The process consists of two stages:

% \textbf{1. Parsing \& Linearization.}

\textbf{(1) Parsing \& Linearization.}
To extract structural features from source code, we utilize Tree-sitter, a high-performance incremental parsing generator capable of building concrete syntax trees for various programming languages. It is particularly robust against syntax errors often found in noisy corpora.
% 待加入参考文献

For a given code snippet $c$, we first generate its AST using Tree-sitter\cite{DBLP:conf/aaai/MouLZWJ16}. 
To reduce the computational complexity of similarity comparisons, we perform a Depth-First Search (DFS) traversal to convert the tree structure into a linearized string sequence in S-expression format\cite{DBLP:journals/pacmpl/AlonZLY19}.
Take the following Python function as an example.

\begin{lstlisting}[language=Python, basicstyle=\ttfamily\small, frame=single, numbers=none]
def add(a, b): return a + b
\end{lstlisting}
After parsing and linearization, its structural representation becomes:
\begin{lstlisting}[basicstyle=\ttfamily\small, breaklines=true, frame=single, numbers=none]
(function_definition (identifier) (parameters (identifier) (identifier)) (return_statement (binary_operator (identifier) (identifier))))
\end{lstlisting}
This representation abstracts away specific variable names (e.g., `a ', `b '), retaining only `identifier ' placeholders and syntactic tags. This allows for a focus on logical structure rather than naming conventions in computation.

\textbf{(2) Similarity Calculation \& Scoring.}
Once the linearized AST sequences are obtained, we quantify the structural similarity using sequence matching. Specifically, for a sample $c_i$ in cluster $C_k$, its structural diversity score $S_{syn}(c_i)$ is defined as the complement of the average sequence similarity to other samples in the cluster:

\begin{equation}
\begin{split}
    S_{syn}(c_i) = 1 - \frac{1}{|C_k|-1}
    \sum_{\substack{c_j \in C_k \\ j \neq i}} 
    \text{SeqSim}(\text{AST}(c_i), \text{AST}(c_j))
\end{split}
\end{equation}

% 需要引用
Here, $\text{SeqSim}(\cdot)$ computes the ratio of the Longest Common Subsequence (LCS)~\cite{DBLP:journals/cacm/Hirschberg75} between two linearized AST strings, implemented via the \texttt{difflib.SequenceMatcher} algorithm. Where we define $S_{syn}(c_i) = 0$ (or 1) if $|C_k|=1$. A higher $S_{syn}$ indicates that the sample possesses a unique syntactic structure within its local cluster, thereby offering richer training signals.

\subsubsection{\textbf{Semantic Diversity ($S_{sem}$)}}
Code Search tasks typically involve complex semantic matching scenarios, ranging from intuitive keyword matching to deep intent understanding. Merely retaining high-similarity samples might ensure alignment accuracy but could lead the model to learn only simple pattern matching (Easy Positives), ignoring deeper or ambiguous semantic associations. To enable the retriever to comprehensively learn semantic hierarchical features and improve generalization across queries of varying difficulty, the training data should retain a complete semantic spectrum from strong to weak alignment. By covering samples from high, medium, and low similarity intervals, the model is forced to learn robust decision boundaries rather than memorizing high-frequency co-occurrences.

We measure the cosine similarity between the code embedding $\mathbf{v}_{c}$ and query embedding $\mathbf{v}_{q}$.
\begin{equation}
    S_{sem}(c_i, q_i) = \cos(\mathbf{v}_{c_i}, \mathbf{v}_{q_i})
\end{equation}
Driven by this rationale, we employ a Stratified Sampling strategy. Instead of strictly maximizing similarity, we sample proportionally from high, medium, and low similarity tiers. This ensures the dataset includes standard positives for basic learning, hard positives for robustness, and edge cases for diversity.

% \subsubsection{Metric 3: Density and Diversity Estimation}
\subsubsection{\textbf{Distributional Representation ($S_{dist}$)}}

In the vector space, data representativeness is a multifaceted concept: it requires covering both the core regions of the feature space (representativeness) and the boundary regions (diversity). Focusing solely on cluster centers leads to mode collapse, while focusing purely on outliers introduces noise. 
To address this trade-off, we introduce a composite metric, denoted as $S_{dist}(\mathbf{v}_i)$, which operates directly on the dense semantic vectors $\mathbf{v}_i$ generated in the semantic clustering stage.
We define the distributional score as a tuple of two complementary sub-metrics:
\begin{equation}
    S_{dist}(\mathbf{v}_i) = (S_{den}(\mathbf{v}_i), S_{div}(\mathbf{v}_i))
\end{equation}
where $S_{den}$ and $S_{div}$ are calculated as follows:

\textbf{1. Representative Density ($S_{den}$).} 
Based on Kernel Density Estimation (KDE), we calculate a density score $S_{den}$ for a sample $\mathbf{v}_i$ within a cluster $C_k$ (size $n$) using a Gaussian kernel $K_h$:
\begin{equation}
    S_{den}(\mathbf{v}_i) = \frac{1}{n} \sum_{j=1}^{n} K_h(\|\mathbf{v}_i - \mathbf{v}_j\|)
    \label{eq:s_den}
\end{equation}
High-density samples are selected as they represent the centroid or "mainstream" patterns of the cluster.

\textbf{2. Coverage Diversity ($S_{div}$).} 
To cover semantic boundaries, we calculate the diversity score $S_{div}$ based on the Euclidean distance to the nearest neighbor:
\begin{equation}
    S_{div}(\mathbf{v}_i) = \min_{j \neq i} \| \mathbf{v}_i - \mathbf{v}_j \|_2
\end{equation}
We select samples with the highest $S_{div}$ scores (i.e., those furthest from neighbors) to maximize the semantic breadth covered by the pruned dataset.

\subsection{Data Filtering and Sampling}
\label{sec:pruning}

After evaluating samples across multiple dimensions, simply aggregating the top-scoring candidates may still result in a dataset plagued by redundancy or engineering incompatibilities. To address this, MAP4CS employs a rigorous ``Filter-then-Rank'' cascade pipeline, combining deterministic hard rules with adaptive soft scoring.

\subsubsection{Step 1: Rule-Based Filtering}
Before performing expensive ranking operations, it is computationally efficient to discard samples that are fundamentally invalid or redundant. This step ensures that the final dataset meets basic engineering standards for efficient LLM training.

We implement a three-stage sequential filter:
\begin{enumerate}
    \item \textbf{Cryptographic Deduplication:} To eliminate exact redundancy, we compute the \textbf{SHA-256 hash} of the normalized code content. Duplicate instances share the same hash and are removed, ensuring each code sample is unique.
    \item \textbf{Context-Window Constraint:} Training retrievers on excessively long sequences leads to truncation and semantic loss. We utilize a pre-trained tokenizer to filter out code snippets exceeding the context window (specifically, $L > 1024$ tokens).
    \item \textbf{Heuristic-based Query Filtering:} Inspired by the syntactic filter in NLQF~\cite{DBLP:conf/icse/SunLL0022}, we directly apply its heuristic rules to cleanse the dataset. Noisy queries containing excessive non-ASCII characters, URL links, or specific metadata tags create noisy labels that hinder retrieval performance. Therefore, we filter out these elements as well as sequences that are too short to convey meaningful intent to ensure semantic clarity.

\end{enumerate}

\subsubsection{Step 2: Weighted Ranking and Sampling}

Theoretical analysis suggests that different quality dimensions contribute differently to the training signal. Structural and semantic consistencies are foundational for correctness, while distributional metrics provide auxiliary regularization. A uniform summation would ignore these varying importances.

For the survivors of the hard filtering, we first perform Min-Max Normalization on their raw scores to map them into a unified range $[0, 1]$. We then compute a composite score $S_{total}$ using a weighted fusion strategy:

\begin{equation}
\begin{aligned}
    S_{total}(i) ={}& w_{syn} \cdot \hat{S}_{syn}(i)
    + w_{sem} \cdot \hat{S}_{sem}(i) 
    + w_{den} \cdot \hat{S}_{den}(i)
    + w_{div} \cdot \hat{S}_{div}(i)
\end{aligned}
\end{equation}

Based on empirical tuning and domain knowledge, we assign weights as follows: $w_{syn} = w_{sem} = 1/3$ to prioritize foundational data correctness, and $w_{den} = w_{div} = 1/6$ to balance distributional representativeness.
Finally, we rank all samples by $S_{total}$ and select the top $\mathcal{K}$ instances (e.g., top 5\%) to form the final pruning dataset $\mathcal{S}^*$.

%% file: sections/5_Result/5_Result.tex
\section{Evaluation}
\label{sec:result}

% 然后需要补充 一个implementation，小结，放在实验开题，说明方法实现的细节，比如一些阈值、聚类算法的实现之类。

\input{sections/5_Result/setup}
\input{sections/5_Result/RQ1}

\input{sections/5_Result/RQ2}
\input{sections/5_Result/RQ3}

%% file: sections/5_Result/setup.tex
To comprehensively evaluate the effectiveness and efficiency of \ourapproach, we design experiments to answer the following three research questions:

\begin{itemize}
    \item \textbf{RQ1 (Effectiveness Analysis):} How does \ourapproach perform on downstream code retrieval tasks? 
    \item \textbf{RQ2 (Ablation Study):} What is the contribution of each component in our framework?
    \item \textbf{RQ3 (Parameter Analysis):} How does the pruning ratio impact model performance? 
\end{itemize}

\subsection{Experimental Setup}

\subsubsection{\textbf{Evaluation Benchmarks.}} 
To assess the generalizability of the fine-tuned models, we evaluate them on four diverse benchmarks using the COIR framework. We categorize these benchmarks into two groups to test different capabilities: \textbf{CodeSearchNet} and \textbf{CoSQA} serve as in-domain datasets to evaluate standard code search performance. Conversely, \textbf{APPS} and \textbf{Synthetic Text2SQL}, which involve retrieving corresponding code or SQL solutions based on problem descriptions, are treated as external retrieval datasets to assess the model's out-of-domain (OOD) generalization.

\textbf{\csn (CSN) \cite{DBLP:journals/corr/abs-1909-09436}:} A widely used dataset in the field of code search, covering six programming languages: Python, Java, Go, PHP, Ruby, and JavaScript. Derived from open-source projects on GitHub, the data consists of (Docstring, Code) pairs, where the query is the function's docstring and the code is a complete function body.

\textbf{\cosqa \cite{DBLP:conf/acl/HuangTSG0J0D20}:} A dataset containing 20,604 annotated pairs of natural language queries and codes. The queries originate from real user search logs on Microsoft Bing, while the codes are Python functions from the CodeSearchNet dataset. Each pair is annotated by at least 3 human annotators, providing high-quality supervision for query-code matching.

\textbf{\apps \cite{DBLP:conf/nips/HendrycksBKMAGB21}:} A challenging benchmark for code generation, measuring the ability of models to take arbitrary natural language specifications and generate satisfactory Python code. It includes 10,000 problems ranging from simple one-line solutions to substantial algorithmic challenges. 

\textbf{\texttwosql \cite{gretel-synthetic-text-to-sql-2024}:} A rich dataset of high-quality synthetic Text-to-SQL samples. It includes approximately 106k records covering 100 distinct domains and a wide range of SQL complexity levels, from simple retrieval to complex joins and aggregations.

\subsubsection{\textbf{Models}}

We conduct experiments using \rev{three} representative pre-trained models: \textbf{\unixcoder}~\cite{DBLP:conf/acl/GuoLDW0022}, an encoder-only architecture widely adopted in the code search domain; \textbf{\gte}~\cite{DBLP:conf/emnlp/ZhangZLXDTLYXHZ24}, a contemporaneous model that has achieved state-of-the-art (SOTA) results across multiple general retrieval benchmarks; and \rev{\textbf{Qwen3-Embedding}~\cite{qwen3embedding2025}, a foundation-model-based dense encoder.} \rev{(Additionally, a widely adopted sparse lexical baseline, \textbf{BM25}~\cite{DBLP:journals/ftir/RobertsonZ09}, is included for baseline comparison in our evaluation).}

% We conduct experiments using two representative pre-trained encoders.
% 其中UniXcoder在code search领域被广泛使用，而gte发表时间相近，且在多个通用检索领域取得sota。

\textbf{\unixcoder:} A unified cross-modal pre-trained model for programming languages. 
    To enhance semantic representation, UniXcoder incorporates multi-modal contents such as code comments and ASTs, where the ASTs are flattened via a one-to-one mapping. 
    The model is pre-trained using multi-modal contrastive learning and cross-modal generation tasks to effectively align representations among programming languages and natural language.

\textbf{\gte:} A general-purpose text embedding model based on the BERT architecture \cite{DBLP:conf/naacl/DevlinCLT19}, trained via multi-stage contrastive learning \cite{DBLP:conf/emnlp/ConneauKSBB17}. 
    Despite treating code merely as unstructured text, GTE achieves superior performance on code retrieval benchmarks by leveraging robust semantic representations learned from massive text pairs.

\rev{\textbf{Qwen3-Embedding}~\cite{qwen3embedding2025}: A state-of-the-art foundation-model-based dense encoder built upon the Qwen3 language model family. In this work, we employ the 0.6B variant, which features a 32k context length and supports instruction-aware text embedding. Notably, Qwen3-Embedding is extensively trained on over 100 languages, including various programming languages, providing robust cross-lingual and code retrieval capabilities.}

\subsubsection{\textbf{Implementation}}
\label{sec:setup:implementation}
% 这些超参数的设置有没有支撑?
% 部分参考他人的工作，部分实验尝试得到的参数。
% We implement all experiments using PyTorch and the Sentence-Transformers framework, maintaining a consistent configuration to ensure fair comparisons. The models are fine-tuned for 6 epochs with a batch size of 4, utilizing the AdamW optimizer with an initial learning rate of 2e-5 and a linear warmup of 10\%. To optimize semantic alignment, we employ the Multiple Negatives Ranking Loss (MNRL). The maximum sequence length is set to 512 tokens. For reproducibility, we fix the random seed at 42 across all runs. Our source code and detailed hyperparameters are available at \url{https://anonymous.4open.science/r/MAP4CS-097A}.
We implement all experiments using PyTorch and the Sentence-Transformers framework. To ensure both reproducibility and fair comparison, our hyperparameter settings are determined through a hybrid approach: aligning with established protocols in prior code retrieval studies~\cite{DBLP:conf/acl/GuoLDW0022, DBLP:conf/icse/ShiWGDZHZS23} and conducting preliminary empirical tuning on the validation set to ensure optimal convergence. Specifically, models are fine-tuned for 6 epochs with a batch size of 4. We utilize the AdamW optimizer with an initial learning rate of 2e-5 and a linear warmup of 10%, 
a configuration widely validated for stabilizing transformer fine-tuning. To optimize semantic alignment, we employ the Multiple Negatives Ranking Loss (MNRL), which effectively leverages in-batch negatives to enhance retrieval distinctiveness. The maximum sequence length is set to 512 tokens to accommodate typical code contexts. For reproducibility, we fix the random seed at 42 across all runs. Our source code and detailed hyperparameters are available at \url{https://anonymous.4open.science/r/MAP4CS-097A}.

%% file: sections/5_Result/RQ1.tex
\subsection{RQ1: Effectiveness Analysis}
\label{sec:rq1}

In this section, we first quantitatively evaluate the retrieval performance of \ourapproach against both random sampling and full-dataset fine-tuning across multiple benchmarks. Beyond standard performance metrics, we qualitatively investigate the underlying mechanisms of our approach by visualizing the semantic distribution shifts and analyzing the linguistic diversity of the pruned data.

\subsubsection{Design}
\label{sec:rq1:design}

To investigate the effectiveness of \ourapproach, we fine-tune \rev{three} representative models: \unixcoder, \gte, \rev{and Qwen3-Embedding,} on two distinct datasets: CSN-Java and \queryfourcode. CSN-Java is a subset from the \csn corpus. \queryfourcode is an LLM-annotated dataset specifically constructed to simulate real-world code retrieval scenarios by generating user-style queries rather than using noisy or descriptive docstrings.

We compare our method against two baselines, Random Sampling and Full Data fine-tuning. \rev{To contextualize the neural retrieval results, we also report BM25 as a non-parametric lexical baseline.} To verify the effectiveness, \ourapproach and Random Sampling use the same data budget, 5\%.

In order to evaluate the effectiveness of code search, we use Mean Reciprocal Rank (MRR)~\cite{DBLP:conf/trec/Voorhees99} as an evaluation metric. MRR is widely adopted in information retrieval (IR)~\cite{DBLP:journals/access/HambardeP23} and code search research~\cite{DBLP:journals/tosem/XieLDZW24,DBLP:conf/icse/ShiWGDZHZS23} to assess ranking quality. It measures the mean of the reciprocal ranks of the first relevant code snippet across all queries.
\begin{equation}
\label{eq:mrr}
\text{MRR} = \frac{1}{N} \sum_{i=1}^{N} \frac{1}{\text{rank}_i}
\end{equation}

We follow the same experimental configuration detailed in Section \ref{sec:setup:implementation}, fine-tuning \unixcoder, \gte, \rev{and Qwen3-Embedding} on both datasets with the strict 5\% data budget.

\input{table/rq1}
\subsubsection{Results}
\label{sec:rq1:results}

Table \ref{tab:rq1_main} presents the retrieval performance across four downstream tasks using different fine-tuning strategies. The results provide compelling evidence for the effectiveness of our proposed framework compared to both random sampling and full-dataset training.

\noindent\textbf{\rev{Comparison with BM25.}}
\rev{BM25 achieves an average score of 0.120. Although it remains competitive on CodeSearchNet (0.360), its performance drops sharply on CoSQA (0.086) and the OOD tasks APPS (0.009) and Text2SQL (0.025). This gap indicates that lexical matching alone is insufficient for capturing user intent and cross-domain semantic alignment in code retrieval. In contrast, dense retrievers generally provide stronger semantic matching, and \ourapproach further improves their fine-tuning efficiency by selecting high-quality training pairs.}

% 加个小标题，这个应该是和随机采样比较
\noindent\textbf{Comparison with Random Sampling.}
Under a strict data budget, \ourapproach consistently outperforms the random sampling baseline across all training datasets and models. 
Specifically, on the CSN-Java dataset with the UniXcoder model, our method improves the average score from 0.166 (Random 5\%) to 0.219, achieving a relative improvement of \textbf{31.9\%}. Similarly, on the \gte model, we observe a substantial gain from 0.113 to 0.178 (\textbf{+57.5\%}). \rev{The same trend holds for Qwen3-Embedding: \ourapproach improves the average score from 0.088 to 0.128 on CSN-Java (\textbf{+45.5\%}) and from 0.069 to 0.102 on Query4Code (\textbf{+47.8\%}).} This demonstrates that our multi-dimensional clustering and filtering strategy successfully identifies high-value core samples, whereas random sampling often captures redundant or unrepresentative data.

% 标题比较合适是训练数据量比较，或者和全集进行比较，然后在里面或总结处再出来这个结论The ``Less is More'' Phenomenon.
% \noindent\textbf{The ``Less is More'' Phenomenon.}
\noindent\textbf{Comparison with Full Data Fine-tuning.}
A striking observation is that our method often matches or even exceeds the performance of fine-tuning on the full dataset (100\% data). On the CSN-Java dataset, training with the full dataset yields surprisingly poor results (e.g., Avg 0.053 for \gte), performing even worse than random sampling. This suggests the original corpus contains severe noise or low-quality pairs that induce negative transfer. By effectively pruning this noise, our method achieves a significantly higher score (0.178), validating its capability as a robust data cleaner.
\rev{This phenomenon is particularly clear when comparing CSN-Java and Query4Code. CSN-Java is derived from raw GitHub repositories and therefore contains more noise, including comment-code mismatches, duplicated snippets, and boilerplate implementations. Consequently, pruning CSN-Java with \ourapproach yields pronounced gains by removing harmful or redundant samples. In contrast, Query4Code is a retrieval-optimized query-code dataset with higher semantic consistency and fewer expression errors. Since its full-data baseline is already strong, there is less room for large absolute improvements.}
On the \queryfourcode dataset, where the data quality is relatively stable, our method still achieves comparable or slightly better performance than full-data fine-tuning (e.g., 0.200 vs. 0.192 on UniXcoder). 
This confirms that our framework can reduce training costs by over 90\% without compromising retrieval accuracy.
\rev{Therefore, the key value of \ourapproach is not to substantially surpass full-data fine-tuning on every already high-quality dataset, but to achieve comparable or better retrieval accuracy under an extremely low training budget.}
Such a result effectively validates the \textbf{``Less is More''} phenomenon, demonstrating that a small, high-quality subset can serve as a superior training signal compared to a massive but noisy corpus.

\noindent\textbf{Robustness Across Models.}
The improvements are consistent across the code-specialized (\unixcoder) and general embedding (\gte) models\rev{, extending even to the stronger foundation-model-based Qwen3-Embedding}. Notably, on the domain-specific \csn evaluation metric, our method achieves a peak score of 0.613 (CSN-Java/\unixcoder), significantly outperforming the pre-trained baseline (0.431) and other fine-tuning strategies.
Crucially, on OOD tasks, our method consistently maintains a performance advantage over baselines. For instance, with \gte trained on Query4Code, \ourapproach improves performance on APPS by \textbf{125\%} compared to random sampling (0.027 vs. 0.012). \rev{For Qwen3-Embedding, \ourapproach also improves every downstream task over both random sampling and full-data fine-tuning under the same training dataset, demonstrating that the proposed pruning strategy remains useful even when the backbone encoder has stronger pre-trained semantic representations. Meanwhile, the competitive zero-shot performance of Qwen3-Embedding suggests that overly noisy or mismatched fine-tuning data can still hurt strong foundation encoders, further motivating quality-aware pruning.} This suggests that by maximizing structural and semantic diversity, \ourapproach helps the model learn generalized representations rather than overfitting to specific domain shortcuts.

% 数据集有In domain，o domain，有个分析维度和结论一个体现这个吧

\finding{\ourapproach demonstrates superior effectiveness in code retrieval tasks by outperforming random sampling baselines and achieving comparable or better results than full-dataset fine-tuning, despite using only 5\% of the training data. Moreover, whether in-domain or OOD, our method consistently outperforms random sampling. This highlights the framework's capability in filtering noise and identifying high-value core samples.}

\subsubsection{Visualization of Semantic Distribution}
\label{sec:Vis}

\begin{figure*}[t]
    \centering
    % Random
    \begin{minipage}[t]{0.32\textwidth}
        \centering
        \includegraphics[width=\linewidth]{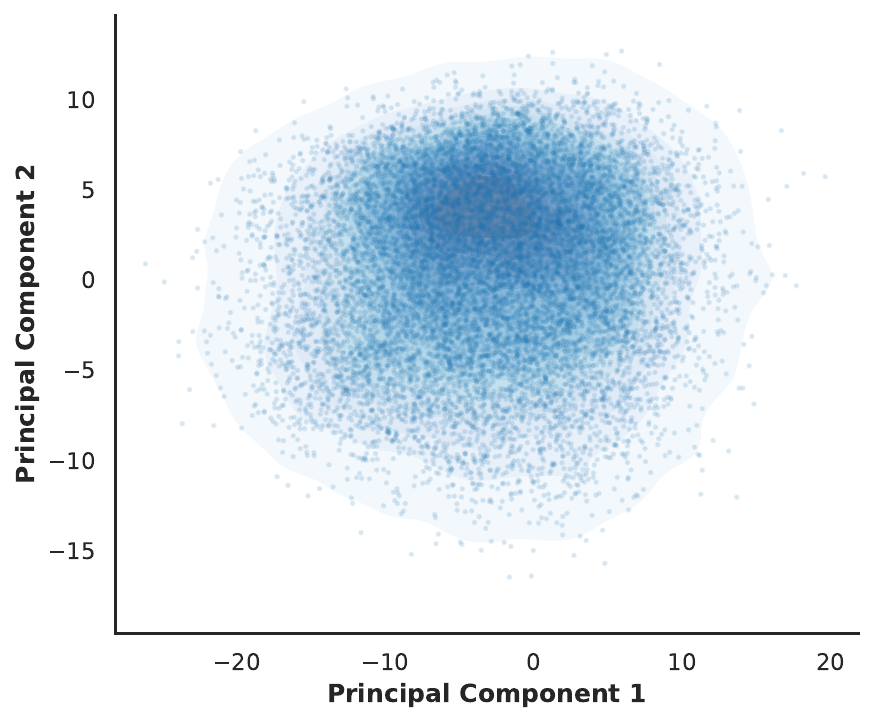}
        \centerline{\small (a) Random Sampling (5\%)}
    \end{minipage}
    \hfill
    % Pruned
    \begin{minipage}[t]{0.32\textwidth}
        \centering
        \includegraphics[width=\linewidth]{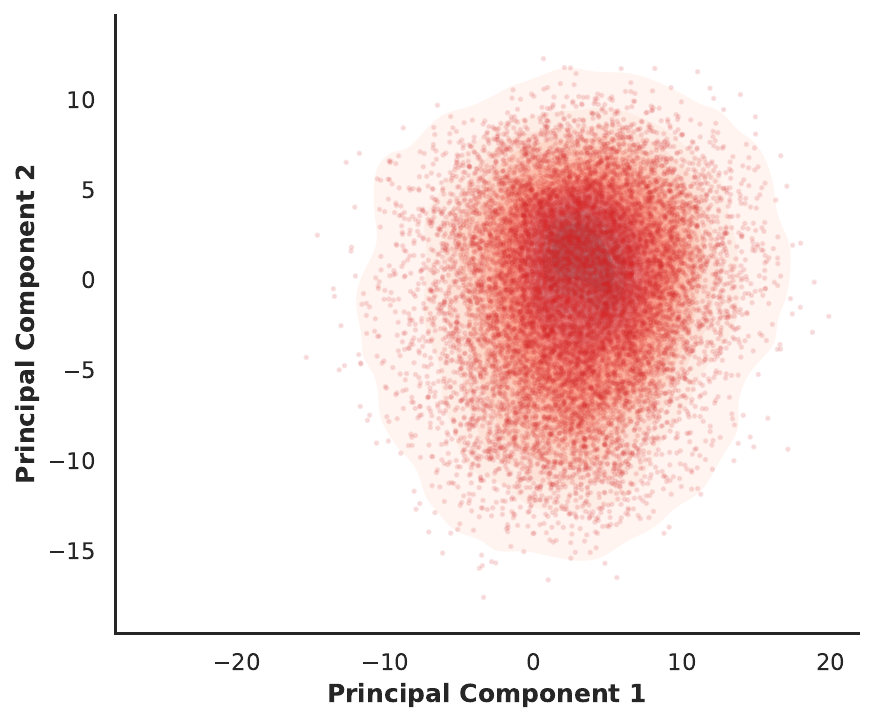} 
        \centerline{\small (b) \ourapproach (5\%)} 
    \end{minipage}
    \hfill
    % Overlay
    \begin{minipage}[t]{0.32\textwidth}
        \centering
        \includegraphics[width=\linewidth]{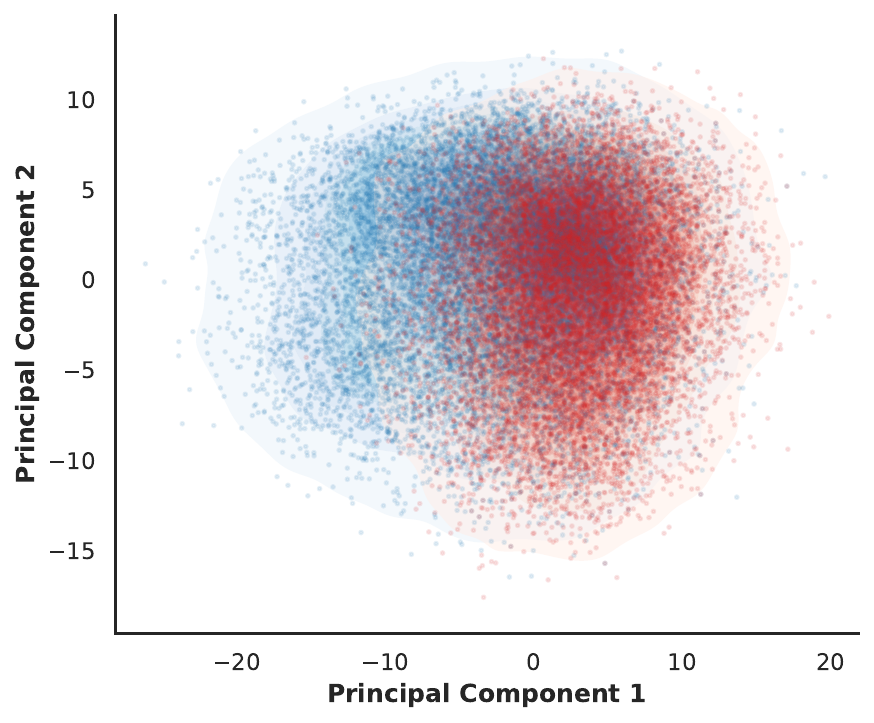} 
        \centerline{\small (c) Overlay Comparison} 
    \end{minipage}
    % \vspace{-5pt}
    \caption{Visualization of semantic distribution in the Code-Query Semantic Space. The Random Sampling baseline (Blue) exhibits a diffuse distribution with scattered outliers, indicating high-entropy noise. In contrast, \ourapproach (Red) demonstrates a more refined distribution, concentrating on high-value semantic regions.}
    \label{fig:pca_vis}
    % \vspace{-10pt}
\end{figure*}

To visually verify the effectiveness of \ourapproach in filtering data, we compare the semantic distribution of our pruned subset against the Random Sampling baseline. 
Specifically, we employ the \unixcoder encoder to extract high-dimensional embeddings from the two subsets generated by \ourapproach and Random Sampling, respectively. These embeddings, derived from the mean pooling of the last hidden state, are then projected into a 2D space using Principal Component Analysis (PCA)~\cite{pearson1901liii}. To ensure a fair comparison within the same semantic manifold, we concatenate the vectors from both groups before fitting the PCA model. Furthermore, to address the overplotting issue in scatter plots and to quantify distributional differences, we visualize the distributions using Kernel Density Estimation (KDE) contours, as shown in Figure \ref{fig:pca_vis}. 
Simply put, KDE helps visualize data density by smoothing discrete points into continuous contours, allowing us to clearly identify the high-density ``core'' regions of the semantic space (consistent with Eq. \ref{eq:s_den}).
% KDE is a non-parametric technique that estimates the probability density function of a random variable, effectively smoothing discrete data points to reveal the continuous underlying structure of the semantic space, computed via Eq. \ref{eq:s_den}.

To quantify the divergence between the distributions, we calculate the \textbf{Jensen-Shannon (JS) Distance}~\cite{DBLP:journals/tit/Lin91} and the \textbf{Overlap Coefficient}~\cite{vijaymeena2016survey}. The Overlap Coefficient measures the overlapping area of two probability distributions (range [0, 1], where higher values indicate greater similarity), while the JS Distance measures the difference based on KL divergence (range [0, 1], where lower values indicate greater similarity). Analysis reveals distinct characteristics in the semantic space.
In the \textbf{Code-Query Semantic Space}, \ourapproach\ exhibits a significantly different distribution profile compared to random sampling. The Overlap Coefficient drops to 0.5549, while the JS Distance increases to 0.4895. This divergence supports two key mechanisms:

\begin{itemize}
    \item \textbf{Effective Denoising:} The random sampling distribution (Blue in Figure \ref{fig:pca_vis}) is characterized by a diffuse spread, mixing high-quality data with a large volume of noisy samples where the code is mismatched or semantically ambiguous relative to the query.
    
    % 为什么能说我们方法覆盖的就是高价值区域，需要更加详细有逻辑、依据的说明
    \item \textbf{Semantic Refinement:} \ourapproach\ (Red) forms a distinct, compact cluster within the semantic space. 
    Notably, our clusters appear more compact, yet the fine-tuned model performance is superior. This suggests that unlike Random Sampling which captures a diffuse spread of data including outliers (potential noise), \ourapproach may effectively isolate and retain samples with high semantic density, proving that data quality—defined by semantic consistency—is more critical than mere coverage for training efficient retrievers.
    
\end{itemize}

% 总结和finding重复了。
% In summary, the significant reduction in the overlap coefficient within the joint code-query space reveals a deeper optimization mechanism: we successfully filter out samples with poor implementation quality or weak semantic alignment. This strategy significantly enhances the quality of code implementations in the final dataset, achieving a ``coarse-to-fine'' refinement.

% The quantified divergence in JS Distance (0.4895) further corroborates that our method creates a distinct training distribution profile, shifting focus from quantity to quality.

\finding{Quantitative analysis of semantic distribution confirms that \ourapproach\ achieves a ``coarse-to-fine'' refinement. By shifting the data distribution away from noisy, mismatched regions (low overlap with random baseline) and concentrating on high-consistency semantic clusters, it constructs a significantly purer and more representative training corpus.}

\subsubsection{Reason Analysis}
\label{sec:rq1:reason}

To investigate the underlying reasons for performance improvement, we analyze the linguistic characteristics of the training data on the \queryfourcode dataset. 

\begin{figure*}[t]
    % \vspace{-0.7em}
    \centering
    % Random
    \begin{minipage}[t]{0.48\textwidth}
        \centering
        \includegraphics[width=\linewidth]{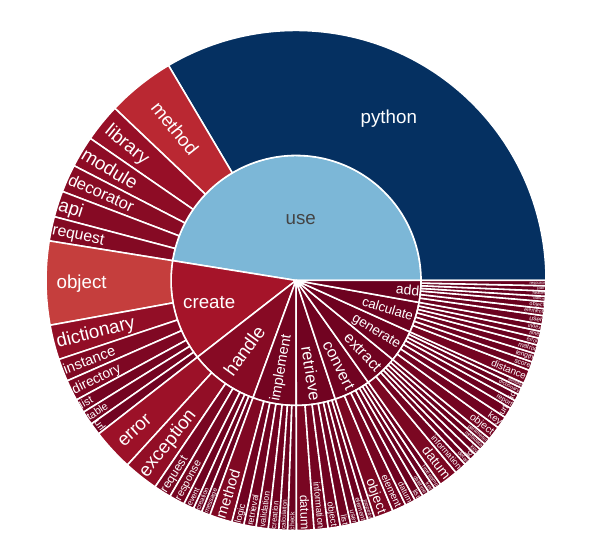}
        \centerline{\small (a) Random Sampling (5\%)}
        % \caption{Semantic Intent Distribution \\ (Random Sampling)}
        % \label{fig:sunburst_random}
    \end{minipage}
    \hfill
    % Pruned
    \begin{minipage}[t]{0.48\textwidth}
        \centering
        \includegraphics[width=\linewidth]{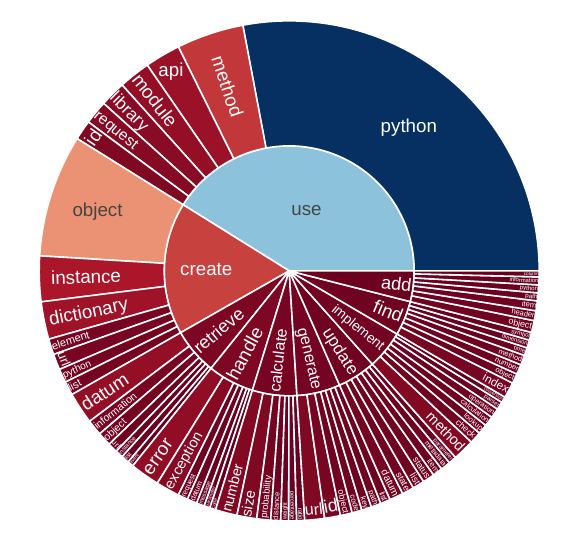} 
        \centerline{\small (b) \ourapproach (5\%)} % 手动写子标题
        % 假设这里有对应的 Pruned 图片

    \end{minipage}
    % \vspace{-5pt}
    \caption{Sunburst plot of the top 10 most frequent verbs, with their corresponding top 7 root nouns within the query subsets obtained via Random Sampling and \ourapproach.}
    \label{fig:sunburst}
    % \vspace{-15pt}
\end{figure*}

First, we visualize the semantic intent distribution by depicting the most common verbs and their corresponding objects within the query subsets obtained via Random Sampling and \ourapproach (Figure \ref{fig:sunburst}).
The analysis reveals that while there is some overlap in high-frequency verbs (e.g., ``use", ``add") between random sampling and our method, the pruned dataset exhibits a cleaner and more focused distribution of object terms, reducing long-tail noise.

In addition, we utilize quantitative metrics to measure the quality of the data. On one hand, our method increases the Type-Token Ratio (TTR) from 0.0402 (Random Sampling) to 0.0579 (\ourapproach). Since TTR represents the ratio of unique words to total words, this substantial increase indicates that \ourapproach effectively performs ``de-duplication'' and ``quality enhancement,'' filtering out repetitive boilerplate code. On the other hand, the Shannon Entropy slightly decreases from 6.269 (Random Sampling) to 6.160 (\ourapproach), a drop of approximately 1.7\%. While entropy represents uncertainty, high entropy in random sampling often stems from noise and long-tail confusion. Our clustering-based approach acts as a denoising mechanism, making the data distribution more representative.

\finding{Compared to random sampling, \ourapproach effectively filters out high-entropy noise while simultaneously achieving higher lexical diversity, resulting in an information-dense training set.}

%% file: table/rq1.tex
\begin{table*}[t]
    \centering
    \caption{Performance comparison on downstream code retrieval tasks.}
    \label{tab:rq1_main}
    \resizebox{0.95\textwidth}{!}{
    \begin{tabular}{ll|cccc|c}
    \toprule

        % --- New Header Row ---
    \multirow{2}{*}{\textbf{Training Dataset}} & \multirow{2}{*}{\textbf{Method}} & \multicolumn{2}{c}{\textbf{In-Domain}} & \multicolumn{2}{c|}{\textbf{Out-of-Distribution}} & \multirow{2}{*}{\textbf{Avg.}} \\
    \cmidrule(lr){3-4} \cmidrule(lr){5-6} % Adding lines to group columns
     & & \textbf{CodeSearchNet} & \textbf{CoSQA} & \textbf{APPS} & \textbf{Text2SQL} & \\
    % ----------------------
    % \textbf{Training Dataset} & \textbf{Method} & \textbf{CodeSearchNet} & \textbf{CoSQA} & \textbf{APPS} & \textbf{Text2SQL} & \textbf{Avg.} \\
    \midrule
    \multicolumn{7}{c}{\cellcolor{gray!10}\rev{\textbf{Lexical Baseline}}} \\
    \midrule
    \textit{None} & \rev{BM25} & \rev{0.360} & \rev{0.086} & \rev{0.009} & \rev{0.025} & \rev{0.120} \\
    \midrule
    \multicolumn{7}{c}{\cellcolor{gray!10}\textbf{\unixcoder}} \\
    \midrule
    \textit{None} & Zero-Shot & 0.431 & 0.122 & 0.007 & 0.124 & 0.171 \\
    \midrule
    \multirow{3}{*}{CSN-Java} 
      & Full Data (100\%)    & 0.365 & 0.092 & 0.010 & 0.030 & 0.124 \\
      & Random Sampling (5\%) & 0.467 & 0.106 & 0.019 & 0.070 & 0.166 \\
      & \textbf{\ourapproach(5\%)}    & \textbf{0.613} & \textbf{0.150} & \textbf{0.036} & \textbf{0.076} & \textbf{0.219} \\
    \midrule
    \multirow{3}{*}{Query4Code} 
      & Full Data (100\%)    & \textbf{0.577} & 0.140 & 0.018 & 0.033 & 0.192 \\
      & Random Sampling (5\%) & 0.542 & 0.156 & 0.010 & 0.032 & 0.185 \\
      & \textbf{\ourapproach(5\%)}    & 0.573 & \textbf{0.160} & \textbf{0.028} & \textbf{0.037} & \textbf{0.200} \\
    \midrule
    \multicolumn{7}{c}{\cellcolor{gray!10}\textbf{\gte}} \\
    \midrule
    \textit{None} & Zero-Shot & 0.200 & 0.180 & 0.018 & 0.127 & 0.131 \\
    \midrule
    \multirow{3}{*}{CSN-Java} 
      & Full Data (100\%)    & 0.126 & 0.018 & 0.007 & 0.060 & 0.053 \\
      & Random Sampling (5\%) & 0.273 & 0.066 & 0.019 & 0.095 & 0.113 \\
      & \textbf{\ourapproach(5\%)}    & \textbf{0.438} & \textbf{0.132} & \textbf{0.028} & \textbf{0.116} & \textbf{0.178} \\
    \midrule
    \multirow{3}{*}{Query4Code} 
      & Full Data (100\%)    & \textbf{0.447} & 0.134 & 0.018 & 0.043 & 0.160 \\
      & Random Sampling (5\%) & 0.374 & \textbf{0.146} & 0.012 & 0.042 & 0.144 \\
      & \textbf{\ourapproach(5\%)}    & 0.432 & 0.140 & \textbf{0.027} & \textbf{0.044} & \textbf{0.161} \\
    \midrule
    \multicolumn{7}{c}{\cellcolor{gray!10}\textbf{\rev{Qwen3-Embedding}}} \\
    \midrule
    \textit{None} & \rev{Zero-Shot} & \rev{0.398} & \rev{0.076} & \rev{0.079} & \rev{0.125} & \rev{0.170} \\
    \midrule
    \multirow{3}{*}{\rev{CSN-Java}} 
      & \rev{Full Data (100\%)}    & \rev{0.160} & \rev{0.006} & \rev{0.002} & \rev{0.013} & \rev{0.045} \\
      & \rev{Random Sampling (5\%)} & \rev{0.260} & \rev{0.014} & \rev{0.020} & \rev{0.058} & \rev{0.088} \\
      & \textbf{\rev{\ourapproach(5\%)}}    & \textbf{\rev{0.337}} & \textbf{\rev{0.066}} & \textbf{\rev{0.036}} & \textbf{\rev{0.073}} & \textbf{\rev{0.128}} \\
    \midrule
    \multirow{3}{*}{\rev{Query4Code}} 
      & \rev{Full Data (100\%)}    & \rev{0.190} & \rev{0.030} & \rev{0.004} & \rev{0.023} & \rev{0.062} \\
      & \rev{Random Sampling (5\%)} & \rev{0.210} & \rev{0.022} & \rev{0.009} & \rev{0.034} & \rev{0.069} \\
      & \textbf{\rev{\ourapproach(5\%)}}    & \textbf{\rev{0.298}} & \textbf{\rev{0.040}} & \textbf{\rev{0.011}} & \textbf{\rev{0.059}} & \textbf{\rev{0.102}} \\
    \bottomrule
    \end{tabular}
    }
    % \vspace{-10pt}
\end{table*}

%% file: sections/5_Result/RQ2.tex
% \vspace{-10pt}
\subsection{RQ2: Ablation Study}
\label{sec:rq2}

In this section, we dissect the contribution of each component within \ourapproach\ to understand the source of its effectiveness.
\subsubsection{Design}
To isolate the impact of different pruning stages, we conduct an ablation study using the \gte model on the \queryfourcode dataset. We compare the following settings (Clustering Only and \ourapproach) against the Random Sampling fine-tuning baseline. Clustering Only applies random sampling within the clusters generated by Single-pass Clustering to select 5\% of the data.

\input{table/rq2_table}
\subsubsection{Results}
Table \ref{tab:rq2_ablation} summarizes the results of the ablation study.
Clustering Only achieves an average score of 0.152, showing a clear improvement over the Random Sampling baseline (0.144). This confirms that clustering-based sampling captures the semantic skeleton and diversity of the original dataset more effectively than pure random sampling. By ensuring coverage across different semantic clusters, the model avoids overfitting to dominant patterns.

Integrating the multi-dimensional sorting and filtering mechanism (\ourapproach) further boosts the performance to 0.161, demonstrating that quality filtering is as important as diversity. The improvement from Clustering Only (0.152) to \ourapproach (0.161) highlights the value of our scoring mechanism. By filtering out low-quality samples within clusters (e.g., mismatched query-code pairs or low-density outliers), \ourapproach ensures that the model trains on the most representative data. Notably, on the challenging APPS task, our method achieves a score of 0.027, significantly outperforming both Clustering Only (0.018) and Random Sampling (0.012). This indicates that high-quality, denoised training data is crucial for the model to generalize well on complex code search tasks, rather than merely memorizing patterns.

\finding{While semantic clustering provides a strong foundation for data diversity, the multi-dimensional filtering and scoring mechanism is the key driver of performance. It eliminates harmful noise and retains high-quality samples, enabling \ourapproach to surpass both random and clustering-based baselines, especially on complex generalization tasks.}

%% file: table/rq2_table.tex
\begin{table*}[t]
    \centering
    \caption{Ablation study of different components in our framework. The best results among fine-tuned models are highlighted in \textbf{bold}.}
    \label{tab:rq2_ablation}
    \resizebox{0.9\textwidth}{!}{
    \begin{tabular}{l|cccc|c}
    \toprule
% --- Grouped Header Row ---
    % --- Grouped Header Row ---
    \multirow{2}{*}{\textbf{Method}} & 
    \multicolumn{2}{c}{\textbf{In-Domain}} & 
    \multicolumn{2}{c|}{\textbf{Out-of-Distribution}} & \multirow{2}{*}{\textbf{Avg.}} \\
    \cmidrule(r){2-3} \cmidrule(l){4-5} % Lines covering specific columns
     & \textbf{CodeSearchNet} & \textbf{CoSQA} & \textbf{APPS} & \textbf{Text2SQL} & \\
    % --------------------------
    % \textbf{Method} & \textbf{CodeSearchNet} & \textbf{CoSQA} & \textbf{APPS} & \textbf{Text2SQL} & \textbf{Avg.} \\
    \midrule
    \multicolumn{6}{c}{\cellcolor{gray!10}\textbf{\gte (Training Dataset: Query4Code)}} \\
    \midrule
    Random Sampling (5\%) & 0.374 & \textbf{0.146} & 0.012 & 0.042 & 0.144 \\
    \midrule
    Clustering Only (5\%) & 0.401 & \textbf{0.146} & 0.018 & 0.041 & 0.152 \\
    \midrule
    \textbf{\ourapproach (5\%)} & \textbf{0.432} & 0.140 & \textbf{0.027} & \textbf{0.044} & \textbf{0.161} \\
    \bottomrule
    \end{tabular}
    }
    \vspace{-10pt}
\end{table*}

%% file: sections/5_Result/RQ3.tex
\subsection{RQ3: Parameter Analysis}
\label{sec:rq3}

In this section, we investigate the trade-off between data scale and model performance, and further analyze the linguistic properties of the sampled subsets to explain the efficiency of our method.

\subsubsection{Design}
To answer RQ3, we conduct a sensitivity analysis on the \queryfourcode dataset. We vary the sampling ratio from 3\% to 100\% (specifically \{3\%, 5\%, 10\%, 20\%, 50\%, 70\%, 100\%\}) and evaluate the performance of both \unixcoder and \gte. We compare our \ourapproach\ against the Random Sampling baseline to assess data efficiency and robustness.

\subsubsection{Results}
Figure \ref{fig:rq3_performance} illustrates the performance trends.
\ourapproach demonstrates exceptional efficiency in low-data scenarios. Even with as little as 3\% of the training data, our pruned subset consistently outperforms the random sampling baseline. Notably, on UniXcoder, the model trained with just 3\% of our pruned data surpasses the model trained with 50\% randomly sampled data. This validates our framework's ability to prioritize the most information-dense samples.

A counter-intuitive phenomenon observed in the random sampling baseline is the performance volatility—and even degradation—as the data ratio increases (e.g., from 5\% to 70\%). This suggests that blindly adding more data introduces detrimental noise. In contrast, our method maintains a stable and superior performance trajectory, confirming that our multi-dimensional filtering acts as an effective ``firewall'' against long-tail noise.

\finding{\ourapproach achieves peak performance with minimal data ($\sim$10\%) and exhibits superior stability compared to Random Sampling, which may suffer from performance degradation due to noise accumulation as data scale increases.}

\subsubsection{Reason Analysis}
\begin{figure*}[t]
    \centering
    % 图 1: Performance Trend
    \begin{minipage}[t]{0.32\textwidth}
        \centering
        \includegraphics[width=\textwidth]{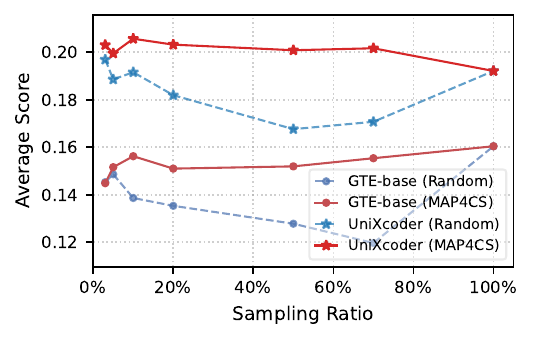}
        \caption{Performance Trend across Sampling Ratios}
        \label{fig:rq3_performance}
    \end{minipage}
    \hfill
    % 图 2: TTR Trend
    \begin{minipage}[t]{0.32\textwidth}
        \centering
        \includegraphics[width=\textwidth]{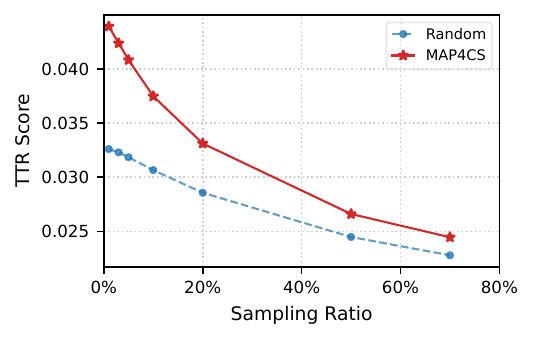}
        \caption{TTR Trend}
        \label{fig:rq3_ttr}
    \end{minipage}
    \hfill
    % 图 3: Entropy Trend
    \begin{minipage}[t]{0.32\textwidth}
        \centering
        \includegraphics[width=\textwidth]{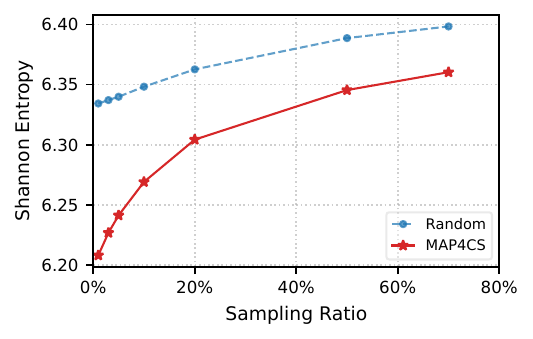}
        \caption{Shannon Entropy Trend}
        \label{fig:rq3_entropy}
    \end{minipage}
    % \vspace{-10pt}
\end{figure*}

To understand why our method achieves superior performance with less data, we analyze the evolution of linguistic quality metrics across different sampling ratios, as shown in Figure \ref{fig:rq3_ttr} and Figure \ref{fig:rq3_entropy}.

% \vspace{-15pt}

Figure \ref{fig:rq3_ttr} reveals a significant ``cold-start'' advantage for our method. In the low-resource regime (Ratio $< 10\%$), the pruned subset maintains a substantially higher TTR compared to random sampling. Although TTR naturally decays as the corpus size grows, the persistent gap highlights that \ourapproach prioritizes non-redundant samples. This implies that the model is exposed to a broader spectrum of lexical and syntactic patterns much earlier in the training process.

As shown in Figure \ref{fig:rq3_entropy}, the random sampling baseline consistently exhibits slightly higher entropy values compared to our pruned subset. While higher entropy typically implies more information, in the context of raw code corpora, it often stems from high-entropy noise. The slightly lower but stable entropy of our method indicates a successful denoising effect. 
% (e.g., arbitrary variable names, garbled text)

By filtering out chaotic, high-entropy elements while enhancing lexical diversity, \ourapproach constructs a more regularized and information-dense training corpus, which explains its superior efficiency.

\finding{Linguistic analysis confirms that the efficiency of \ourapproach stems from its ability to maximize lexical diversity (High TTR) while filtering out high-entropy noise, effectively solving the trade-off between data scale and quality.}

%% file: sections/6_Threats.tex
\section{Threats to Validity}
\label{sec:threats}

\parabf{Internal Validity.}
One potential threat to internal validity is the selection of hyperparameters, which can significantly impact model performance. To mitigate this, we \rev{maintain} consistent hyperparameters for all model fine-tuning experiments, ensuring consistency and fairness in our evaluations.
Additionally, there is a risk of data contamination; pre-trained models like \unixcoder might have been exposed to the \csn corpus during their initial pre-training phase, which could artificially inflate evaluation results on in-domain datasets. To mitigate this concern and verify the true generalization capability of our method, we incorporate APPS and Synthetic Text2SQL as out-of-distribution (OOD) benchmarks. These datasets represent distinct tasks with vastly different data distributions, ensuring that the observed performance gains stem from the high-quality pruned data rather than prior data memorization.

\parabf{External Validity.}
The generalizability of our findings might be limited by the selection of retrieval models and datasets. To mitigate model-specific biases, we evaluate three diverse dense neural architectures: \unixcoder (code-specialized), \gte (general-purpose), and \rev{Qwen3-Embedding (foundation-model-based)}, alongside \rev{BM25 as a sparse lexical baseline}. This broad model coverage minimizes the risk that our findings are biased toward a specific retrieval mechanism.
To ensure robustness across data distributions, we conduct experiments on datasets with varying noise levels and construction paradigms, specifically CSN-Java (raw open-source repositories) and Query4Code (LLM-optimized queries). Furthermore, to address potential bias towards specific languages, we extend our evaluation using the \csn and Synthetic Text2SQL benchmarks, which collectively cover seven programming languages (Go, Java, JavaScript, PHP, Python, Ruby, and SQL). This multi-lingual evaluation demonstrates that the effectiveness of \ourapproach is not confined to a single programming language but is applicable across diverse software ecosystems.

\rev{A final threat to external validity involves the gap between academic benchmarks and real-world industrial code search. Large-scale enterprise search logs are strictly confidential due to trade secrets, proprietary business logic, and sensitive development contexts, making it difficult to fully replicate industrial deployment conditions. Nevertheless, our benchmark suite partially bridges this gap: notably, CoSQA consists of real user queries sampled from Microsoft Bing search logs, effectively capturing realistic developer intents and natural-language query styles. We leave the validation of \ourapproach on anonymized industrial search logs as a direction for future work.}

%% file: sections/7_Related.tex
\section{Related Work}
\label{sec:related}

\subsection{Neural Code Retrieval}

Code retrieval aims to identify the most relevant code snippet from a corpus given a natural language query. Early approaches primarily rely on keyword matching techniques, such as TF-IDF~\cite{DBLP:journals/corr/abs-2001-09896} and BM25~\cite{DBLP:journals/ftir/RobertsonZ09}. With the advent of deep learning, neural code retrieval has become the dominant paradigm~\cite{DBLP:conf/icse/GuZ018}. Many modern approaches employ a bi-encoder architecture, where code and queries are mapped into a shared vector space~\cite{DBLP:conf/icse/GuZ018}.

Pre-trained models like CodeBERT~\cite{DBLP:conf/emnlp/FengGTDFGS0LJZ20}, GraphCodeBERT~\cite{DBLP:conf/iclr/GuoRLFT0ZDSFTDC21}, and UniXcoder~\cite{DBLP:conf/acl/GuoLDW0022} employ masked language modeling and cross-modal tasks to learn robust representations, significantly advancing the state-of-the-art. Building on this, researchers have introduced code-specific optimization strategies: ContraCode~\cite{DBLP:conf/emnlp/0001JZA0S21} leverages compiler-generated code variants for contrastive learning to capture functional semantics; CodeRetriever~\cite{DBLP:conf/emnlp/LiGSQZYQJCD22} enforces code-text alignment through unimodal and bimodal contrastive losses; and CodeT5~\cite{DBLP:conf/emnlp/0034WJH21} adopts identifier-aware pre-training tasks, excelling in both generation and retrieval. Recently, general-purpose text embeddings like GTE~\cite{DBLP:conf/emnlp/ZhangZLXDTLYXHZ24} have also shown promise in code domains.

While these works focus on designing better model architectures or pre-training objectives, they typically train on massive, unfiltered corpora. In contrast, our work focuses on the data-centric perspective, investigating how to efficiently fine-tune these strong backbones using a small but highly representative subset of data.

\subsection{Data Pruning and Selection}

Data pruning aims to select a high-quality subset of training data without compromising model performance. This field has evolved significantly across different domains.

\noindent\textbf{General NLP and LLM Fine-tuning.}
In general NLP, strategies range from heuristic rules (e.g., filtering based on length, perplexity, or stop-word ratios)~\cite{DBLP:conf/lrec/WenzekLCCGJG20} to sophisticated metric-based approaches.
With the rise of Large Language Models (LLMs), the focus has shifted to Instruction Tuning data selection. The ``Less Is More for Alignment'' (LIMA) hypothesis~\cite{DBLP:conf/nips/ZhouLX0SMMEYYZG23} demonstrates that a handful of carefully curated examples can outperform massive datasets. Subsequent works like AlpaGasus~\cite{DBLP:conf/iclr/ChenLYWGYTS0HJ24} and MoDS~\cite{DBLP:journals/corr/abs-2311-15653} leverage powerful teacher models to score and filter data based on quality, complexity, and diversity. Other approaches leverage training dynamics, such as Influence Functions~\cite{DBLP:conf/icml/KohL17} or Forgetting Events~\cite{DBLP:conf/iclr/TonevaSCTBG19}, to identify samples that contribute most to model generalization. However, these methods typically treat data as linear text sequences, ignoring the rigid syntactic structures inherent in programming languages.

\noindent\textbf{Data Selection for General Retrievers.}
In the broader field of information retrieval (e.g., web search, QA), data strategies have primarily focused on data augmentation or negative sampling rather than positive pruning for efficiency. For instance, techniques like ANCE~\cite{DBLP:conf/iclr/XiongXLTLBAO21} and RocketQA~\cite{DBLP:conf/naacl/QuDLLRZDWW21} mine "hard negatives" to sharpen decision boundaries.
While effective, these approaches essentially \textit{increase} the training load to improve performance. Research on pruning \textit{positive} training samples to reduce computational costs remains sparse in the retrieval domain. Moreover, existing general selection methods do not account for the specific semantic gap between code logic and documentation. As a result, directly applying general text pruning metrics to code retrieval often yields sub-optimal results.

\noindent\textbf{Code-Specific Data Cleaning.}
Data cleaning in software engineering has evolved from simple preprocessing to sophisticated quality assurance pipelines.
\textit{Deduplication} is a foundational step; Allamanis~\cite{DBLP:conf/oopsla/Allamanis19} demonstrates that code duplication artificially inflates performance metrics and causes data leakage. Consequently, modern large-scale code datasets, such as The Stack~\cite{DBLP:journals/tmlr/KocetkovLALMJMF23} and StarCoder~\cite{DBLP:journals/tmlr/LiAZMKMMALCLZZW23}, implement aggressive Near-deduplication (e.g., MinHash) and PII redaction as standard protocols.

Beyond deduplication, recent research focuses on filtering low-quality code using \textit{dynamic} or \textit{static} signals.
On the dynamic side, execution feedback is leveraged to verify functional correctness. For instance, Shi et al.~\cite{DBLP:conf/icse/ShiWGDZHZS23} and Han et al.~\cite{DBLP:conf/emnlp/HanKHDL23} harness test case execution results to align code retrieval models with ground-truth correctness, filtering out snippets that fail to compile or pass tests.
However, execution-based methods are computationally expensive and hard to scale.
Alternatively, static approaches rely on heuristic rules (e.g., line length, parser checks) or learned metrics.
Gunasekar et al.~\cite{DBLP:journals/corr/abs-2306-11644} introduce a classifier-based approach to select "textbook-quality" code, significantly boosting model performance with less data.
Similarly, embedding-based pruning strategies~\cite{DBLP:conf/nips/SorscherGSGM22} have been explored to remove outliers in the latent space.

Despite these advancements, existing code pruning methods often rely on \textit{single-dimensional} metrics---either focusing solely on syntactic validity, execution outcomes, or distinctive semantic features~\cite{DBLP:conf/lrec/WenzekLCCGJG20,DBLP:conf/icml/KohL17}.
Our framework, \textbf{\ourapproach}, addresses this limitation by introducing a \textit{multi-dimensional} perspective.
We integrate semantic consistency, syntactic structure (AST), and distributional density to perform a holistic assessment of data value, specifically tailored for retrieval-augmented generation tasks.

\subsection{Retrieval-Augmented Code Generation (RACG)}
RACG enhances Large Language Models (LLMs) by retrieving relevant context from external knowledge bases, thereby mitigating hallucinations and adapting to private domains. Recently, RACG has been widely applied across various SE tasks. 

% In terms of code completion, ReACC~\cite{DBLP:conf/acl/LuDHGHS22} leverages retrieved similar snippets as additional context, while RepoCoder~\cite{DBLP:conf/emnlp/ZhangCZKLZMLC23} introduces an iterative retrieval-generation loop to capture broader repository-level dependencies.
% For in-context learning, CEDAR~\cite{DBLP:conf/icse/NashidSM23} demonstrates that retrieval-based prompt selection significantly boosts performance in assertion generation and program repair.
% Furthermore, domain-specific applications are emerging. For instance, Vul-RAG~\cite{DBLP:journals/corr/abs-2406-11147} enhances vulnerability detection via knowledge-level retrieval; Codegen4Libs~\cite{DBLP:conf/kbse/LiuYLDWP23} employs a two-stage retrieval approach for library-oriented code generation; and Ou et al.~\cite{DBLP:journals/corr/abs-2503-18305} adopt repository context retrieval to improve code translation.
% Additionally, research on "what to retrieve" is also advancing; for instance, Gu et al.~\cite{DBLP:journals/corr/abs-2503-20589} empirically investigate the impact of retrieval granularity on generation performance.
RACG has been widely adopted across tasks such as code completion~\cite{DBLP:conf/acl/LuDHGHS22,DBLP:conf/emnlp/ZhangCZKLZMLC23}, in-context learning~\cite{DBLP:conf/icse/NashidSM23}, and domain-specific applications including vulnerability detection~\cite{DBLP:journals/corr/abs-2406-11147}, library generation~\cite{DBLP:conf/kbse/LiuYLDWP23}, and translation~\cite{DBLP:journals/corr/abs-2503-18305}. Recent studies also investigate retrieval granularity~\cite{DBLP:journals/corr/abs-2503-20589} to optimize generation.

Despite the diverse applications of RACG, its performance relies heavily on the quality of the retriever. While extensive research focuses on optimizing the generator, expanding applications, or refining inference-time retrieval strategies (e.g., re-ranking)~\cite{DBLP:conf/emnlp/WangLGB0H23}, the efficiency of \textit{adapting} the retriever itself remains under-explored.
Maintaining an effective retriever for rapidly evolving codebases requires frequent updates. Our work addresses the training efficiency bottleneck in RACG by demonstrating that high-performance retrievers can be trained with significantly less data, making continuous adaptation feasible for resource-constrained environments.

%% file: sections/8_conclusion.tex
\section{Conclusion}
\label{sec:conclusion}

In this paper, we address the critical challenge of efficient and robust retriever fine-tuning for RAG in software engineering. We propose \textbf{\ourapproach}, an adaptive multi-dimensional data pruning framework specifically tailored for code search tasks. By selecting a small, high-quality core subset from massive, noisy code corpora, \ourapproach innovatively integrates syntactic structure, semantic consistency, and distributional representation
to evaluate sample value, and employs a rigorous rule-based filtering pipeline to ensure data uniqueness and engineering quality.

Our experiments demonstrate that \ourapproach consistently outperforms random sampling baselines across both in-domain and out-of-domain benchmarks. Remarkably, using only \textbf{$\sim$5\%} of the training data, our method achieves performance comparable or even superior to fine-tuning on the full dataset. 
% Specifically, our ablation study reveals that while semantic clustering ensures diversity (breadth), the structure-aware filtering significantly boosts the model's generalization capability on complex tasks (e.g., APPS).
This empirical evidence strongly supports the ``less is more'' hypothesis in data-centric AI: data quality and density are more critical than sheer volume for effective domain adaptation.

Furthermore, our linguistic analysis reveals an adaptive optimization mechanism inherent in \ourapproach. We find that the framework automatically acts as a \textit{de-duplicator} for redundant corpora and a \textit{denoiser} for chaotic corpora, effectively solving the cold-start problem in low-resource regimes. This dual capability allows it to construct a training corpus that is both lexically diverse and information-dense. In future work, we plan to extend \ourapproach to larger-scale decoder-only LLMs 
% for generative tasks 
and investigate iterative pruning strategies for continuous learning scenarios.

%% file: sections/9_data_availability.tex
\section{Data availability}
To facilitate the replication study, we have released our data and code at: ~\url{https://anonymous.4open.science/r/MAP4CS-097A}.